\documentclass[
reprint,
amsmath,
amssymb,
aps,
prb,
floatfix,
longbibliography,
]{revtex4-2}

\usepackage[
colorlinks = true, 
linkcolor = blue, 
anchorcolor = blue, 
citecolor = blue, 
filecolor = blue, 
urlcolor = blue, 
pdfauthor={author},
unicode, 
bookmarks=false,
pdfpagelabels=false
]{hyperref}

\usepackage{graphicx}
\usepackage{bm}
\usepackage{amsfonts, ascmac, mathtools, braket, multirow, color, xcolor, physics, bbm, tabularx, enumerate, comment}
\usepackage{dcolumn}

\begin{document}

\title{Digital quantum simulation of the Kitaev quantum spin liquid}

\author{Seongjun Park}
\author{Eun-Gook Moon} 
\email{egmoon@kaist.ac.kr}
\affiliation{Department of Physics, Korea Advanced Institute of Science and Technology, Daejeon 34141, Republic of Korea}

\begin{abstract}
The ground state of the Kitaev quantum spin liquid on a honeycomb lattice is an intriguing many-body state characterized by its topological order and massive entanglement. 
One of the significant issues is to prepare and manipulate the ground state as well as excited states in a quantum simulator. 
Here, we provide a protocol to manipulate the Kitaev quantum spin liquid via digital quantum simulation. 
A series of unitary gates for the protocol is explicitly constructed, showing its circuit depth is an order of $O(N)$ with the number of qubits, $N$. 
We demonstrate the efficiency of our protocol on the IBM Heron r2 processor for $N = 8$ and $12$. 
We further validate our theoretical framework through numerical simulations, confirming high-fidelity quantum state control for system sizes up to $N = 450$, and discuss the possible implications of these results.    

\end{abstract}
\maketitle

\section{Introduction}
Quantum spin liquids (QSL) are highly entangled quantum phases that lack magnetic ordering even at zero temperature due to strong quantum fluctuations \cite{Balents_2010,Savary_2016,Zhou_2017,Knolle_2019}.
Among them, the Kitaev quantum spin liquid (KQSL), defined on a two-dimensional (2D) honeycomb lattice, is a prime example of an exactly solvable qubit/spin model hosting fractionalized excitations and topological order \cite{Kitaev_2006}. 
One of the key properties is strong magnetic frustration associated with the anisotropic nature of its spin exchange interactions, where the interaction direction depends explicitly on the bond orientation.
The model hosts both Abelian and non-Abelian anyonic excitations, depending on the anisotropy of spin exchange interactions, whose quasiparticles are of great interest not only from a fundamental physics perspective but also for their potential use in fault-tolerant topological quantum computation \cite{Kitaev_2003,Nayak_2008}.

Since its introduction, the KQSL model has stimulated extensive theoretical, experimental, and numerical investigations, which have provided valuable insights into the physical properties and potential applications of the KQSL phase \cite{Janssen_2016,Gohlke_2018,Bolens_2018,Liang_2018,Zhu_2018,Nasu_2018,Hickey_2019,Yoshitake_2020,Chari_2021,Hwang_2022,Noh_2024}.
However, the experimental realization of the KQSL phases still remains one of the most significant challenges in physics due to their theoretical elegance and potential applications.
Current research efforts are broadly categorized into two complementary directions:
(i) the search for real materials that intrinsically exhibit Kitaev-like interactions, and
(ii) the development of quantum simulation platforms that can realize the Kitaev model in a highly controlled setting.

\begin{figure}[b]
\setcounter{figure}{0}
\includegraphics[width=1.0\linewidth]{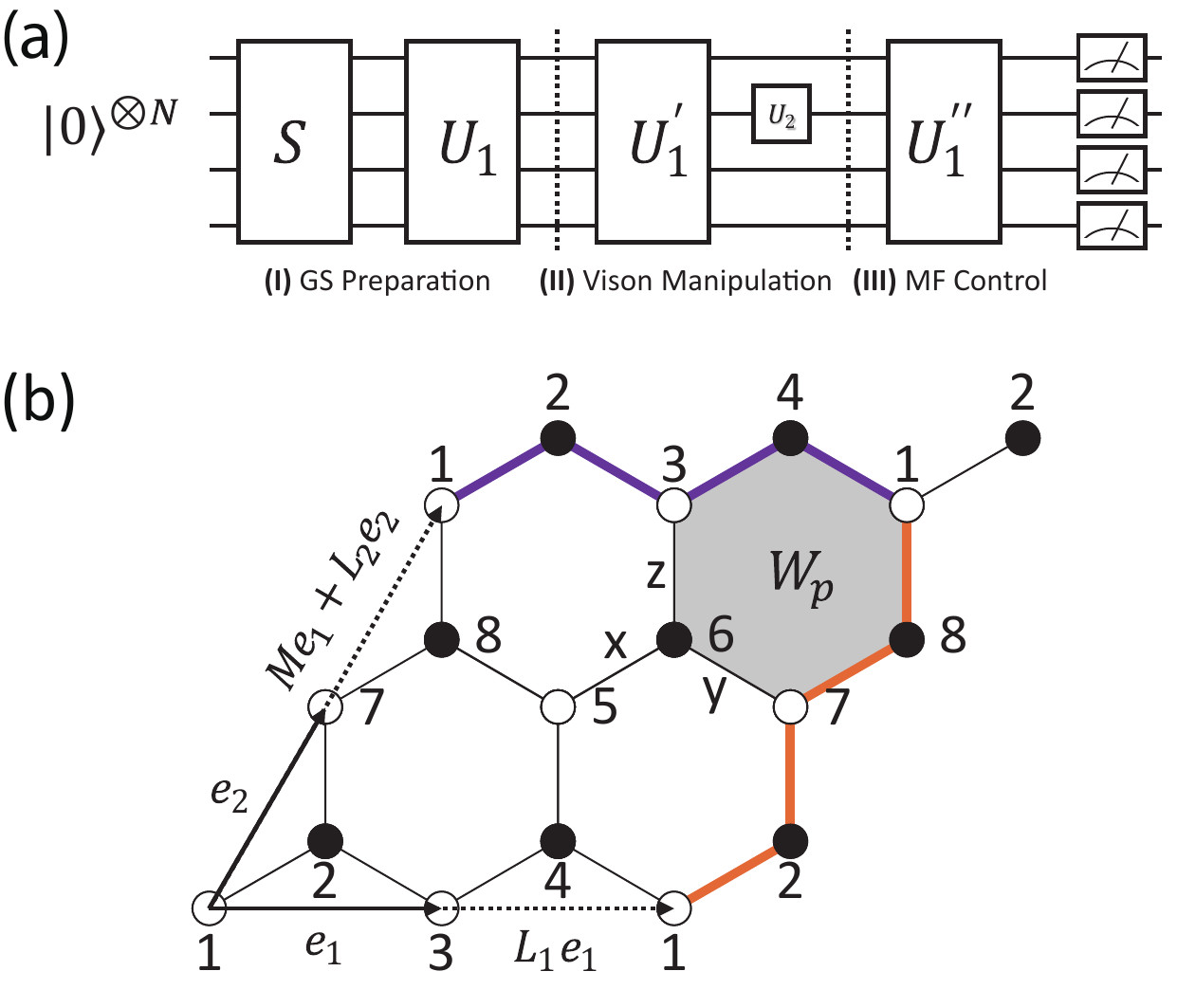}
\caption{(a) Quantum circuit representation of four processes: ground state preparation, vison manipulation, Majorana fermion control, and Majorana fermion readout.
(b) The geometry of Kitaev honeycomb model on the torus, ($L_1=L_2=2$ and $M=0$). $M$ is the twisting parameter of torus. 
Dashed arrows connect the identical sites on the torus.
Purple (orange) loop indicates non-contractible loop $W_{X}$ ($W_{Y}$) on the torus.}
\end{figure}

\begin{figure*}[t]
\setcounter{figure}{1}
\includegraphics[width=1.0\linewidth]{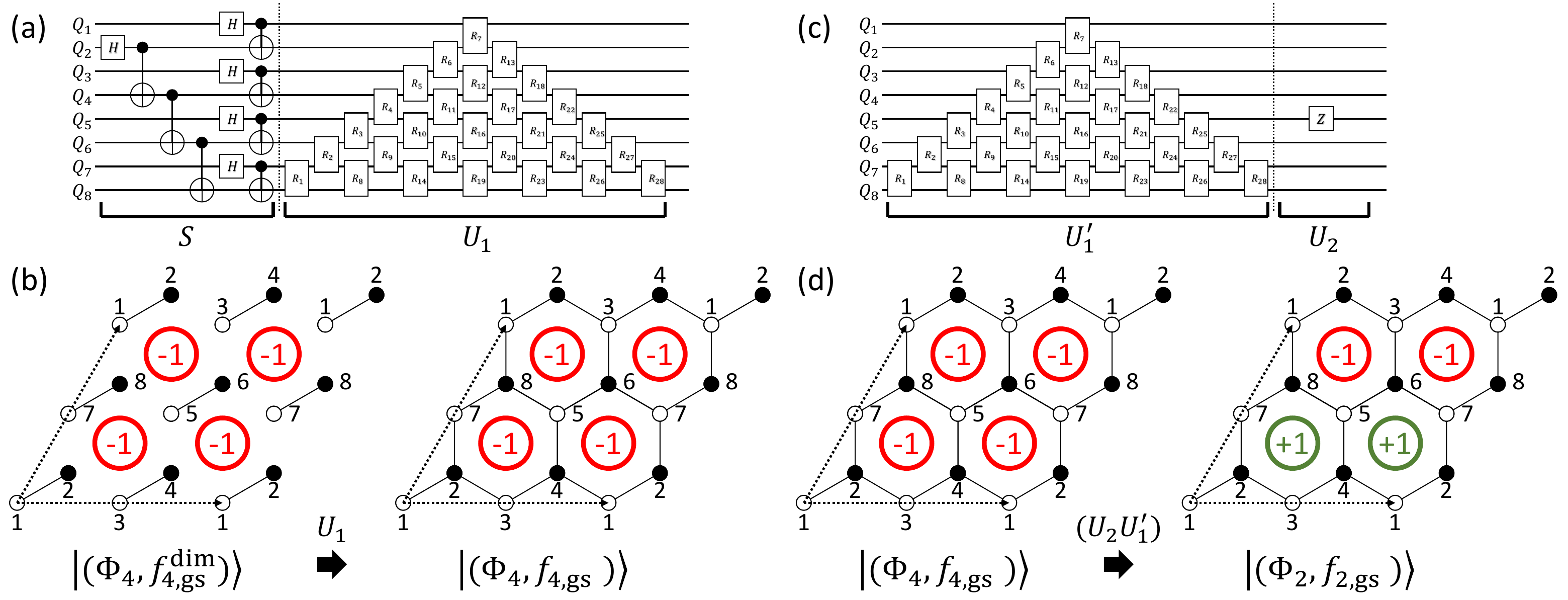}
\caption{(a) Schematic diagram of the quantum circuit for GS preparation. 
The dashed line divides the quantum circuit into two parts: preparation of the state $|(\Phi_{4},f_{4,\text{gs}}^{\text{dim}})\rangle$ ($S$ part), and creation of KQSL ground state through $U_{1}$ operator.
(b) Graphical representation of the mapping implemented by $U_{1}$ operator. 
The Hamiltonian $H_{\text{dim}}$ \eqref{Eq:H_dim} only allows the Kitaev interaction in $x$-direction links.
(c) Schematic diagram of the quantum circuit for vison manipulation. 
(d) Graphical representation of the mapping implemented by $U_{2}U_{1}^{\prime}$ operator. 
The vison pair is annihilated from full-vison sector.
}
\end{figure*}

In the first approach, the material-based strategy was initiated by a theoretical proposal \cite{Jackeli_2009} suggesting that spin-orbit coupled Mott insulators could host effective Kitaev interactions through a combination of strong electronic correlations and relativistic spin-orbit effects.
In particular, systems with strong spin-orbit coupling and Coulomb interaction, such as $\alpha\mathrm{-RuCl}_3$ \cite{Plumb_2014,Sandilands_2015,Koitzsch_2016,Yadav_2016, Kim_2016,Winter_2018,Wulferding_2020, Li_2021,Tanaka_2022,Imamura_2023} and various Cobalt-based honeycomb magnets \cite{Viciu_2007, Songvilay_2020,Das_2021, Lin_2021, Takeda_2022,Zhang_2023, Kim_2024}, have emerged as promising candidates.
These materials typically belong to the family of layered transition metal oxides or halides and display substantial anisotropic exchange interactions arising from their crystal structures and electronic configurations \cite{Chaloupka_2010, Plumb_2014, Janssen_2017}.
However, a central difficulty in this approach is that real materials rarely conform to the idealized Kitaev model.
They often feature competing interactions, including isotropic Heisenberg exchange and off-diagonal couplings, which complicate the identification of a pure KQSL phase \cite{Rau_2016,Trebst_2022}.
In spite of recent dramatic advances in experiments, the manipulation and associated identification of KQSL states calls for future research efforts in the material-based approach.

The second approach involves the use of quantum simulators, which offer an alternative pathway by engineering the Kitaev Hamiltonian in synthetic systems where parameters can be precisely tuned.
Several promising platforms have been explored for this purpose, including trapped ions \cite{Schmied_2011}, ultracold atoms in optical lattices \cite{Sun_2023}, arrays of Rydberg atoms \cite{Kalinowski_2023,Chen_2024}, superconducting circuits \cite{You_2010,Kells_2014,Sameti_2019}, superconducting qubits \cite{Will_2025}, and networks of quantum dots \cite{Cookmeyer_2024}.
Each of these platforms offers distinct advantages in terms of controllability, scalability, and the ability to measure quantum correlations directly.
In parallel, progress has been made on the algorithmic side, where variational quantum algorithms such as the variational quantum eigensolver (VQE) have been employed to approximate the ground state of the KQSL \cite{Jahin_2022,Bespalova_2021,Li_2023,Kalinowski_2023}.
These algorithms leverage the structure of near-term quantum processors to efficiently explore the large Hilbert space of the model.

Despite the remarkable progress, significant challenges remain, particularly when it comes to the preparation, control, and measurement of the exotic quasiparticle excitations that define the KQSL phase.
Namely, the KQSL supports fractionalized excitations—visons and itinerant Majorana fermions—that require careful manipulation and detection strategies in stark contrast to conventional excitations such as magnons.
In particular, the ability to create and braid non-Abelian anyons is essential to realize topological quantum gates, but this remains technically demanding on current platforms especially when the number of qubits increases.

In this paper, we present a protocol aimed at addressing these challenges by proposing concrete strategies for preparing, manipulating, and measuring two key quasiparticles of the KQSL—the vison and the Majorana fermion—on a programmable quantum simulator.
Our approach is based on a digital quantum simulation that leverages the structure of the KQSL model and is compatible with existing gate-based quantum hardware.
Recent studies have demonstrated using digital quantum simulation to prepare and probe nontrivial many-body states \cite{Kokail_2019, Smith_2019, Satzinger_2021}.
Specifically, our strategy draws inspiration from recent advances in the control of non-Abelian anyons in related spin liquid models \cite{Xu_2023,Lensky_2023,Andersen_2023,Iqbal_2024,Minev_2024} and aims to extend those techniques to the KQSL setting.
We emphasize not only the theoretical formulation of these protocols but also their practical implementation potential on quantum devices such as IBM Heron r2 processor.

\section{Kitaev model and Protocol}
\subsection{Two quasiparticle excitations}

The Kitaev honeycomb model is composed of qubits (spin-1/2 particles) on a honeycomb lattice whose Hamiltonian is 
\begin{equation}
H_{\text{K}}=-\sum_{\langle i,j \rangle}J_{ij}^{\alpha}\sigma_{i}^{\alpha}\sigma_{j}^{\alpha}-\sum_{(i,j,k)}K_{ijk}\sigma_{i}^{x}\sigma_{j}^{y}\sigma_{k}^{z}.
\label{Eq:H_Kitaev_0}
\end{equation}
The first term, called the Kitaev interaction, indicates direction-dependent Ising interactions between nearest-neighbor spins, and the second term is three spin interactions resulting from an effective magnetic field. We follow the index convention introduced in Kitaev's original paper \cite{Kitaev_2006}, imposing the torus geometry.

The model is exactly solvable with the vison ($Z_2$ flux) operator on a single plaquette, as shown in Fig. 1(b),
\begin{gather*}
W_{p}=\sigma_{4}^{z}\sigma_{1}^{x}\sigma_{8}^{y}\sigma_{7}^{z}\sigma_{6}^{x}\sigma_{3}^{y}.
\end{gather*}
The presence (absence) of a vison on plaquette $p$ corresponds to $W_{p}=-1$ ($W_{p}=+1$). 
The $Z_2$ flux operators commute with the Hamiltonian, and with each other ($[H, W_{p}]=0$ and $[W_{p}, W_{p'}]=0$), the entire Hilbert space is split into a set of subspaces characterized by a vison configuration. 
Within the subspace, the original Hamiltonian becomes a non-interacting itinerant Majorana fermion model with a background vison configuration \cite{Kitaev_2006}, which is also discussed in Appendix A to be self-contained. 
Thus, the information on itinerant Majorana fermions and visons is necessary to construct eigenstates, and the notation $|(\Phi, f)\rangle$ is used to characterize visons ($\Phi$) and itinerant Majorana fermions ($f$). 
\begin{figure}[t]
\setcounter{figure}{2}
\includegraphics[width=1.0\linewidth]{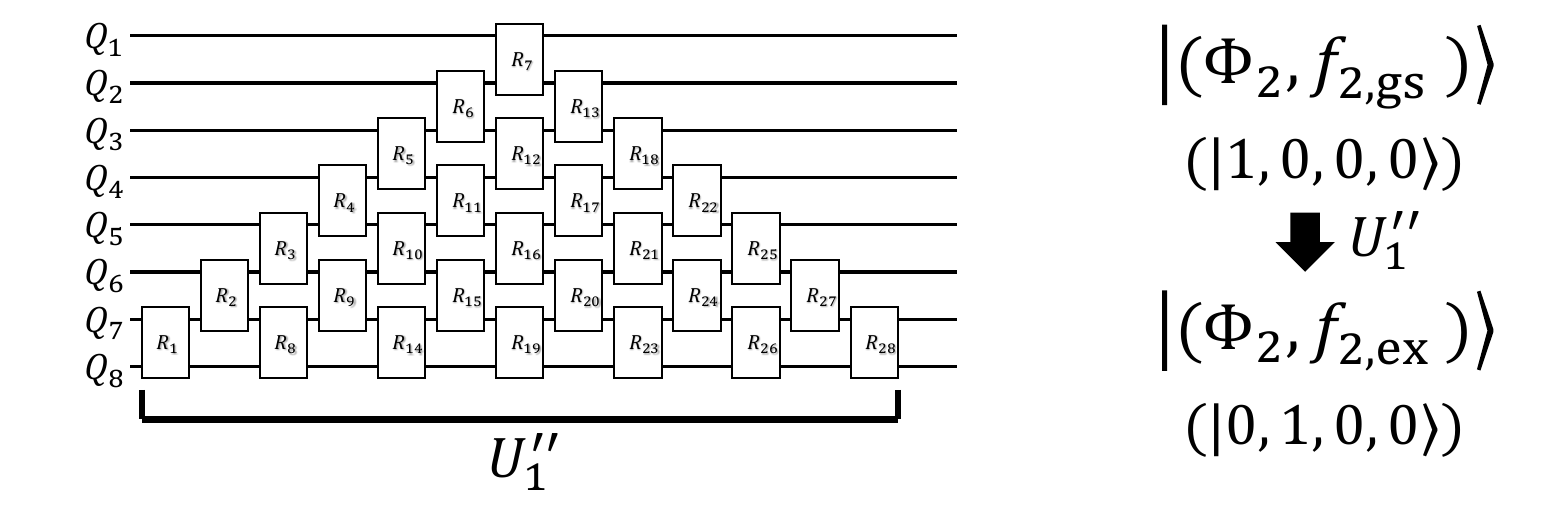}
\caption{Schematic diagram of the quantum circuit for Majorana fermion control.
The $U_{1}^{\prime\prime}$ operator annihilates the lowest mode fermion and creates the second lowest mode fermion.}
\end{figure}

\subsection{Protocols for digital quantum simulation}
Two different eigenstates of the KQSL Hamiltonian \eqref{Eq:H_Kitaev_0} may be formally written as 
\begin{equation}
|(\Phi_{\text{f}},f_{{\text{f}}})\rangle=U_{2}(\Phi_{\text{f}}, \Phi_{\text{i}}) U_{1}(\Phi_{\text{i}} ;  f_{{\text{f}}},  f_{{\text{i}}} ) |(\Phi_{\text{i}},f_{{\text{i}}})\rangle,
\end{equation}
introducing two types of unitary operators, $U_{1}$ and $U_{2}$. 
As manifested in the notation,  the $U_{1}$ operator gives unitary rotation of Majorana fermions within the initial vison sector $\Phi_{\text{i}}$ and the $U_{2}$ operator is for the change of vison configurations from $\Phi_{\text{i}}$ to $\Phi_{\text{f}}$. 
The $U_{2}$ operator may be written as a string of the Pauli matrix, 
\begin{equation}
\label{Eq:U_2_B}
U_{2}(\Phi_{\text{f}}, \Phi_{\text{i}} )=\sigma_{i_{n}}^{\alpha_{n}}\dotsc\sigma_{i_{2}}^{\alpha_{2}}\sigma_{i_{1}}^{\alpha_{1}} ,
\end{equation}
to change the vison configurations.
For the same vison configurations, ($\Phi_{\text{i}}=\Phi_{\text{f}}$), one can use the identity operator. 
Note that the form of a $U_{2}$ operator is not unique whose form determines the form of $U_{1}$ operator.

The Majorana fermion rotation operator $U_1$ may be written as 
 \begin{equation}
\label{Eq:U_1_B}
U_{1}=e^{-iH(B)},\;-iH(B)=\frac{1}{4}\sum_{j,k}{B}_{jk}c_{j}c_{k},
\end{equation}
where $c_{j}$ is a Majorana fermion operator on a site $j$ and $B$ is the ($N\times{}N$) real skew-symmetric matrix. 
The skew-symmetric matrix is determined by $(\Phi_{\text{i}}, \Phi_{\text{f}})$, $(f_{{\text{i}}}, f_{{\text{f}}})$, and $Z_2$ gauge field notation. 
Note that the form of $U_{1}$ operator is identical to part of the variational ansatz introduced in Ref. \cite{Jahin_2022} except the fact that we map one itinerant Majorana fermion with $Z_2$ gauge field to one qubit while the Jordan-Wigner transformation was used to map two itinerant Majorana fermions to one qubit.
 
Our protocol to prepare and manipulate KQSL states consists of three steps with the trivial initial state, $| \Psi_0 \rangle \equiv \prod_j \otimes | 0 \rangle_j$, illustrated in Fig. 1(a). 
\begin{itemize}
\item Step 1: GS preparation,  $| \Psi_1 \rangle \equiv (U_1 S)  | \Psi_0 \rangle $.
\item Step 2: Vison manipulation,  $| \Psi_2 \rangle \equiv (U_2 U_1^{'})  | \Psi_1 \rangle $.
\item Step 3: Majorana Fermion control, $| \Psi_3 \rangle \equiv U_1^{''} |\Psi_2 \rangle $.
\end{itemize}
Combining the three steps enables access to an arbitrary eigenstate of the KQSL Hamiltonian.
Three Majorana fermion rotation operators ($U_1$, $U_1^{'}$, and $U_1^{''}$) and one vison configuration change operator ($U_2$) are used. 
Below, we illustrate how the unitary operator is constructed for the each process, in the context of the eight-qubit model, referring to Appendix B for systems with an arbitrary number of qubits.

One of the key parts of our protocol is to construct the $U_{1}$ operator \eqref{Eq:U_1_B} in the fermion space and transform it into a form that can be implemented on a quantum circuit.
Our main strategy is to express the $U_1$ operator as a sequential product of $R$ operations \eqref{Eq:R_operation}; the $R$ operations can be implemented using Clifford gates in combination with the RZ gate.
\begin{gather*}
    U_{1}=\prod_{j=1}^{M} \otimes R(n_{j},\theta_{j}), \quad R(n,\theta)=e^{-\frac{i}{2}(\theta{}u_{n,n+1})\sigma_{n}^{\alpha}\sigma_{n+1}^{\alpha}}.
\end{gather*}
Specifically, for a system with eight ($N$) qubits, a $U_1$ operator can be written as a sequential product of 28 ($N(N-1)/2$) $R$ operations, with the link direction index $\alpha\in\{\text{$x$, $y$, $z$}\}$ and a gauge configuration $u_{n,n+1}$.
 {Decomposing the $U_1$ operator into a sequence of $R$ operations enables its execution on a quantum circuit, requiring a circuit depth of $O(N)$, while the total number of $R$ operations scales as $O(N^2)$.}

\begin{figure*}[t]
\setcounter{figure}{3}
\includegraphics[width=1.0\linewidth]{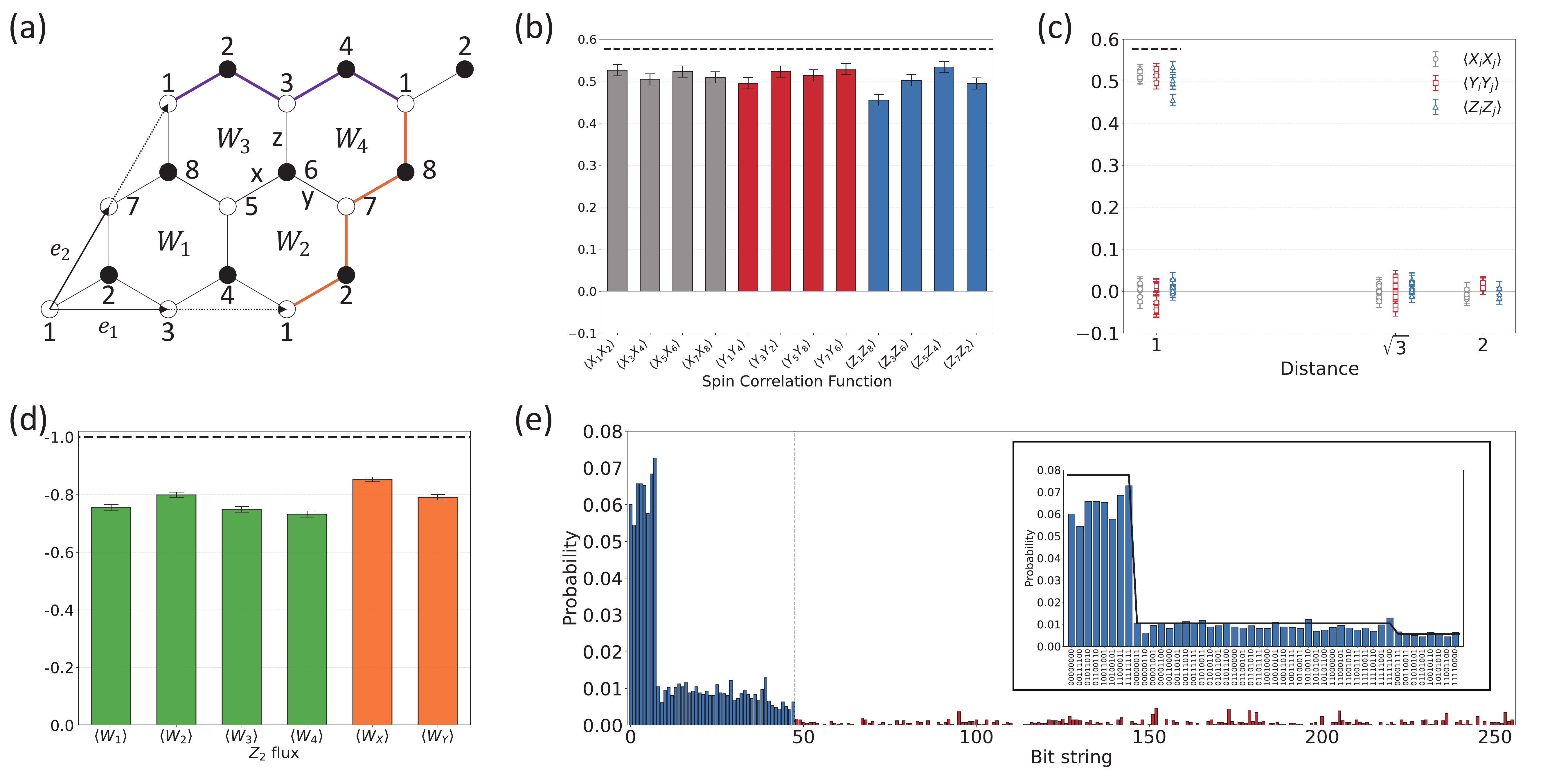}
\caption{Ground state preparation. (a) Geometry of KQSL on the torus. 
Dashed arrow connects identical sites. 
Purple (orange) loop indicates non-contractible loop $W_{X}=-\sigma_{1}^{z}\sigma_{2}^{z}\sigma_{3}^{z}\sigma_{4}^{z}$ ($W_{Y}=-\sigma_{1}^{y}\sigma_{2}^{y}\sigma_{7}^{y}\sigma_{8}^{y}$) on torus.
(b) The spin correlation obtained from the data set (4096 shots in total). 
The spin correlation ($\langle\sigma^{\alpha}_{i}\sigma^{\alpha}_{j}\rangle$) is measured for 12 links connecting the pair of nearest neighbors on the torus, with specific $\alpha$. 
The dashed line indicates the spin correlation value obtained from theory.
The color (gray, red, and blue) indicates the $\alpha$=$x$, $y$, and $z$, respectively.
(c) The measured spin correlation function as a function of distance between two sites. 
The spin correlation function is obtained for all 28 pairs on the torus.
(d) The measured expectation value of four $Z_{2}$ flux operator (green) and two Wilson loop operators (orange).
The dashed line indicates the value obtained from theory.
(e) The (quasi) probability distribution over 256 bit strings obtained from data set (4096 shots in total). 
Blue (red) color indicates that theory predicts its nonzero (zero) probability.
(Inset) The probability distribution over 48 bit strings. 
The solid line indicates the theoretically predicted probability distribution.}
\end{figure*}

First, in the ground state preparation process, the $S$ operator is introduced to initialize the vison configuration and Wilson loop variables, replacing the projection operator onto a specific vison sector \cite{Satzinger_2021, Jin_2023,Will_2025}, see Fig. 2(a) for the circuit design.
Starting from the initial state, $| \Psi_0 \rangle \equiv \prod_{j=1}^8 \otimes | 0 \rangle_j$, we create the state $|(\Phi_{4},f_{4,\text{gs}}^{\text{dim}})\rangle$ by applying the Hadamard (H) and controlled NOT (CX) gates to the quantum circuit ($S$).
The resulting state $|(\Phi_4, f_{4,\text{gs}}^{\text{dim}})\rangle$ is the ground state of a Hamiltonian $H_{\text{dim}}$ \eqref{Eq:H_dim}, characterized by a full-vison configuration where all plaquette operators act as $W_p|\psi\rangle = -1|\psi\rangle$ and the nontrivial Wilson loop eigenvalues $W_X|\psi\rangle = W_Y|\psi\rangle = -1|\psi\rangle$ with
\begin{gather*}
\begin{aligned}
    W_{X}&=-\sigma_{1}^{z}\sigma_{2}^{z}\sigma_{3}^{z}\sigma_{4}^{z},\\
    W_{Y}&=-\sigma_{1}^{y}\sigma_{2}^{y}\sigma_{7}^{y}\sigma_{8}^{y},\\
\end{aligned}
\end{gather*}
where the notation for the site index is presented in Fig. 1(b). 
We then apply the unitary rotation in the fermion space ($U_{1}$) after applying the $S$ operator. 
The $U_{1}$ operator we constructed provides the mapping between the states $|(\Phi_{4},f_{4,\text{gs}}^{\text{dim}})\rangle$ and $|(\Phi_{4},f_{4,\text{gs}})\rangle$, the ground states of respective Hamiltonians $H_{\text{dim}}$ \eqref{Eq:H_dim} and $H_{\text{K}}$, with four-vison configuration, see Fig. 2(b). 
The $U_{1}$ operator provides the mapping from the local fermion modes (Abelian phase) to non-local fermion modes (non-Abelian phase).
 {We stress that the ground state of the eight-qubit KQSL model resides in the four-vison configuration. 
In contrast, the ground state of the KQSL model in larger system sizes corresponds to the zero-vison (vison-free) configuration \cite{Vojta_2015}.}

Second, in the vison manipulation process, we change the vison configuration by applying a set of local spin operators ($U_{2}$).
For example, one can annihilate the vison pair by applying the $U_{2}=\sigma_{5}^{z}$, see Fig. 2(c) for the circuit design.
As it changes the $Z_2$ gauge field, the $U_{2}$ operator rearranges the Majorana fermion \cite{Baskaran_2007}. 
To solve this problem, we apply the $U_{1}^{'}$ operator, prior to applying the $U_2$ operator, to compensate for the change in Majorana fermion affected by the $U_2$ operator.
The $U_{2}U_{1}^{\prime}$ operator we constructed provides the mapping between the states $|(\Phi_{4},f_{4,\text{gs}})\rangle$ and $|(\Phi_{2},f_{2,\text{gs}})\rangle$, the fermionic ground states of different vison sectors, see Fig. 2(d). 

Third, in the Majorana fermion control, we can access the fermionic excited state by applying the $U_{1}^{\prime\prime}$ operator to the state $|(\Phi_{2},f_{2,\text{gs}})\rangle$: this maps the lowest energy fermion-occupied state to the second lowest energy fermion-occupied state, see Fig. 3.
As this process leaves the vison sector unchanged, it is implemented solely as a rotation within the fermion space ($U_1^{''}$).

\begin{figure*}[t]
\setcounter{figure}{4}
\includegraphics[width=1.0\linewidth]{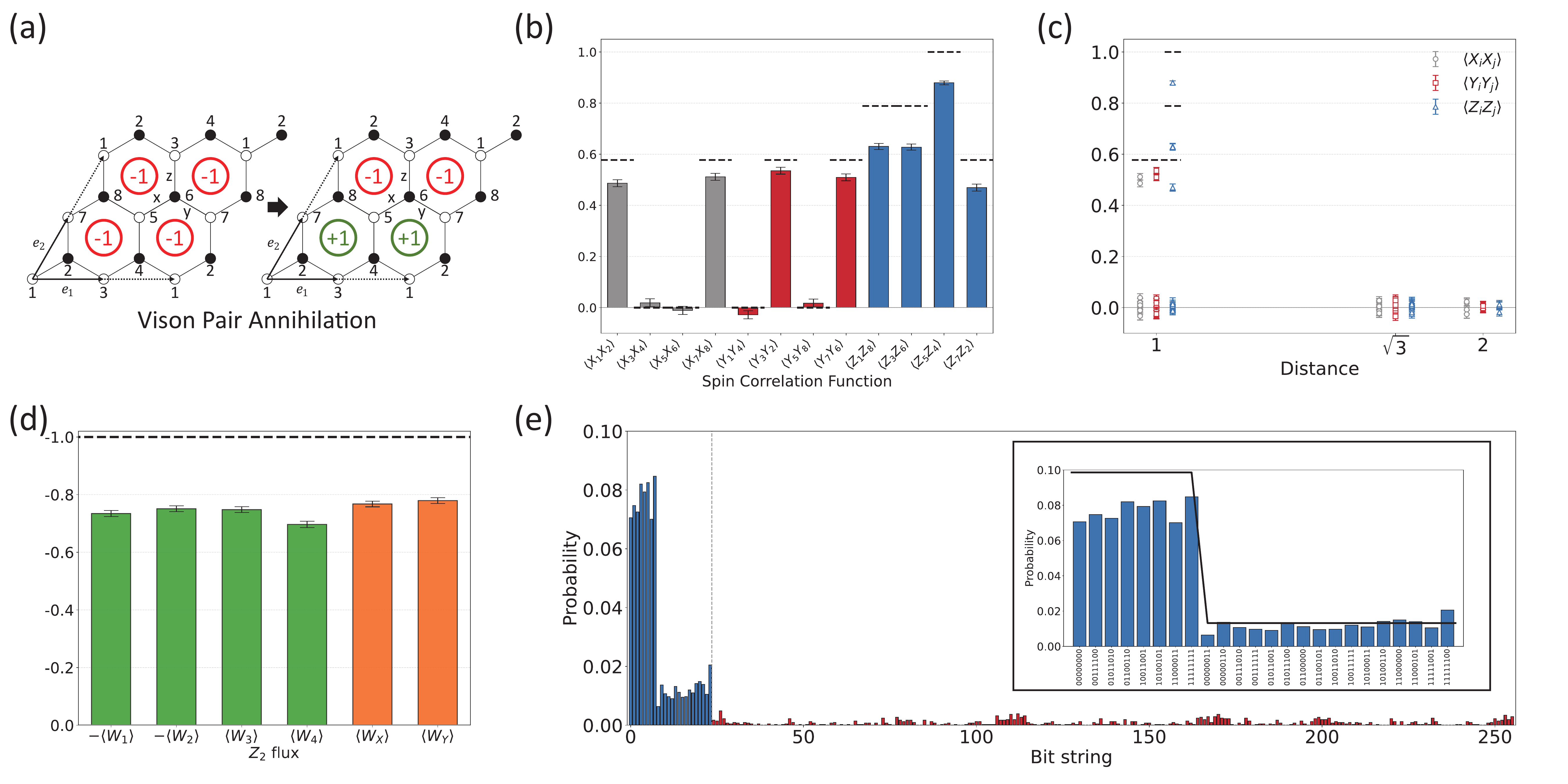}
\caption{Vison manipulation. (a) Graphical representation of the mapping implemented by $U_{2}U_{1}^{\prime}$ operator. 
The vison pair ($W_1$ and $W_2$) is annihilated in this process.
(b) The spin correlation obtained from the data set (4096 shots in total). 
The spin correlation ($\langle\sigma^{\alpha}_{i}\sigma^{\alpha}_{j}\rangle$) is measured for 12 links connecting the pair of nearest neighbors on the torus, with specific $\alpha$. 
The dashed line indicates the spin correlation value obtained from theory.
The theory predicts zero spin correlation for $\langle{}X_{3}X_{4}\rangle$, $\langle{}X_{5}X_{6}\rangle$, $\langle{}Y_{1}Y_{4}\rangle$, and $\langle{}Y_{5}Y_{8}\rangle$.
The color (gray, red, and blue) indicates the $\alpha$=$x$, $y$, and $z$, respectively.
(c) The measured spin correlation function as a function of distance between two sites. 
The spin correlation function is obtained for all 28 pairs on the torus.
(d) The measured expectation value of four $Z_{2}$ flux operator (green) and two Wilson loop operators (orange).
(e) The (quasi) probability distribution over 256 bit strings obtained from data set (4096 shots in total). 
Blue (red) color indicates that theory predicts its nonzero (zero) probability.
(Inset) The probability distribution over 24 bit strings. 
The solid line indicates the theoretically predicted probability distribution.}
\end{figure*}

\section{Digital Quantum Simulation}
Our protocol is applied to a 156-qubit quantum processor, IBM Heron r2 processor "ibm-marrakesh," to realize the ground and excited states of the KQSL Hamiltonian.

A few remarks are as follows. 
First, our simulation targets eigenstates of eight-qubit KQSL model with $K/J=0$, which can be easily generalized.
The index notation for the eight-qubit KQSL model is illustrated in Fig. 4(a). 
We also perform the simulation with the 12-qubit KQSL model whose results are shown and discussed in Appendix C. 
Second, to suppress the experimental noise, we use the dynamical decoupling XY-4, which consists of four $\pi$ pulses applied along alternating axes \cite{Ezzell_2023}.
Lastly, we employed additional strategies to reduce the circuit depth in the digital quantum simulation.
The explicit form of the full quantum circuits for each process is illustrated in Appendix D.
Below, we present our results of digital quantum simulations step-by-step.

\subsection{Ground state preparation}

We verify that the prepared state $|\Psi_1\rangle$ correctly reproduces the ground state of the eight-qubit KQSL model. 
We perform the quantum state tomography and measurement of vison and spin correlation functions.

The measured data shows good agreement with the exact ground state of the eight-qubit KQSL Hamiltonian, as indicated by the comparison with the values  from exact diagonalization calculations.
Figures 4(b) and 4(c) present the measured spin correlation functions. 
Theoretically, the spin correlation function $\langle\sigma^{\alpha}_{i}\sigma^{\alpha}_{j}\rangle$ is nonzero if and only if the link ($i,j$) connects neighboring sites associated with a specific direction $\alpha$. 
Our measurement results identify 12 such nonzero correlations, while the others are strongly suppressed, consistent with theoretical expectations.
We evaluate the energy of the prepared state $|\Psi_1\rangle$ using the measured spin correlation functions. 
The experimentally measured energy is
\[{\langle{}E\rangle}_{\text{exp}}=-6.1083 (\pm0.1613) \,J,\]
which shows reasonably good agreement with the exact ground state energy
\[{\langle{}E\rangle}_{\text{exact}}=-6.9282\,J,\] 
obtained from exact diagonalization.
After applying basis transformations to each qubit, we further measure the vison and Wilson loop operators, as shown in Fig. 4(d).
Finally, Fig. 4(e) shows the (quasi) probability distribution of the prepared state $|\Psi_1\rangle$.

\begin{figure*}[t]
\setcounter{figure}{5}
\includegraphics[width=1.0\linewidth]{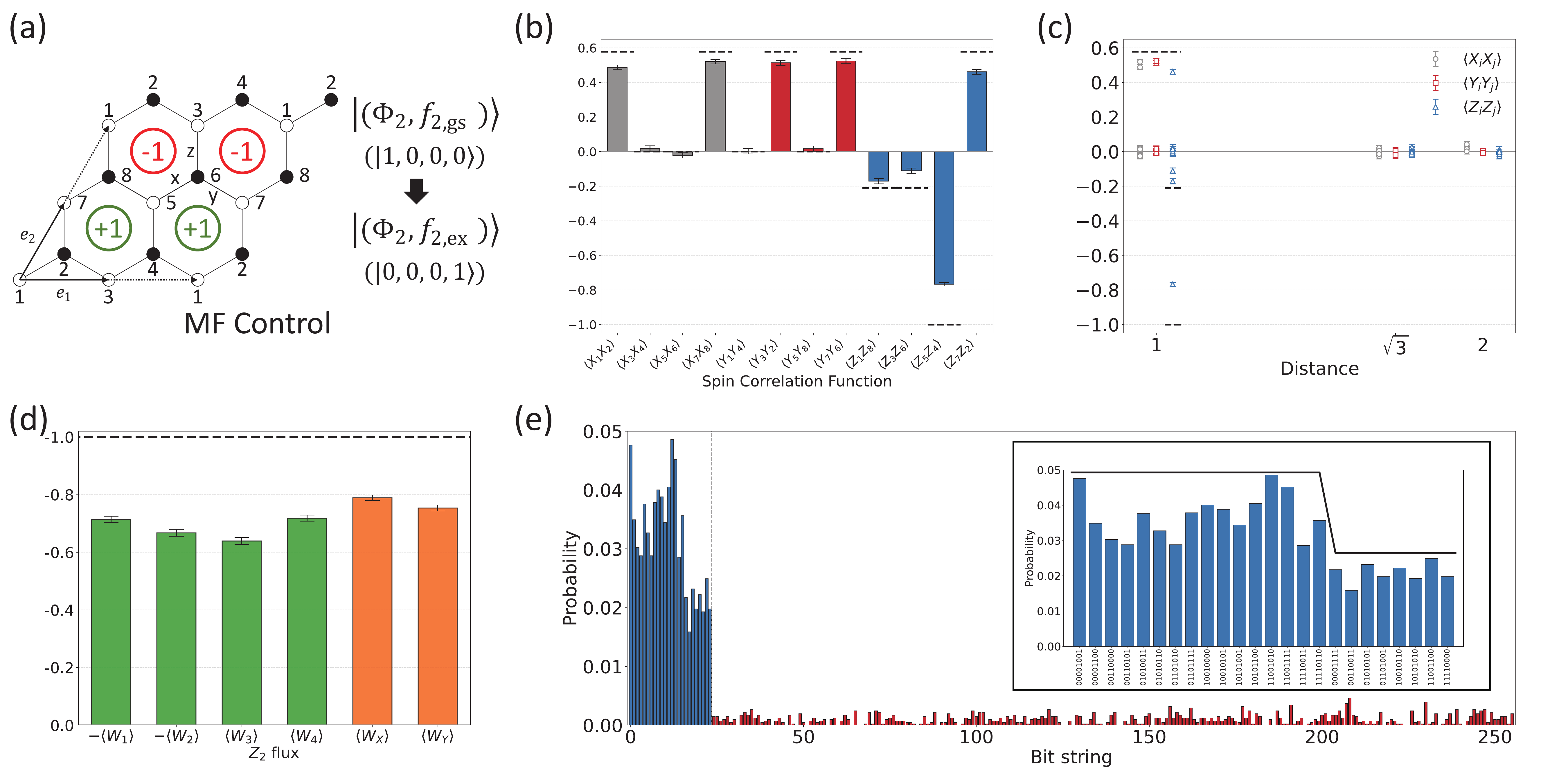}
\caption{Majorana fermion control. (a) Graphical representation of the mapping implemented by $U_{1}^{\prime\prime}$ operator. 
The $U_{1}^{\prime\prime}$ operator annihilates the lowest mode fermion and creates the fourth lowest mode fermion.
(b) The spin correlation obtained from the data set (4096 shots in total). 
The spin correlation ($\langle\sigma^{\alpha}_{i}\sigma^{\alpha}_{j}\rangle$) is measured for 12 links connecting the pair of nearest neighbors on the torus, with specific $\alpha$. 
The dashed line indicates the spin correlation value obtained from theory.
The theory predicts zero spin correlation for $\langle{}X_{3}X_{4}\rangle$, $\langle{}X_{5}X_{6}\rangle$, $\langle{}Y_{1}Y_{4}\rangle$, and $\langle{}Y_{5}Y_{8}\rangle$.
The color (gray, red, and blue) indicates the $\alpha$=$x$, $y$, and $z$, respectively.
(c) The measured spin correlation function as a function of distance between two sites. 
The spin correlation function is obtained for all 28 pairs on the torus.
(d) The measured expectation value of four $Z_{2}$ flux operator (green) and two Wilson loop operators (orange).
(e) The (quasi) probability distribution over 256 bit strings obtained from data set (4096 shots in total). 
Blue (red) color indicates that theory predicts its nonzero (zero) probability.
(Inset) The probability distribution over 24 bit strings. 
The solid line indicates the theoretically predicted probability distribution.
}
\end{figure*}

\subsection{Vison manipulation}

We investigate the vison manipulation process by preparing a new state, $|\Psi_2\rangle$, derived from the eight-qubit KQSL ground state by removing a vison pair ($W_{1}$ and $W_{2}$), as illustrated in Fig. 5(a).

To characterize the resulting state, we repeat the same set of measurements performed in the ground state preparation process. 
As theoretically expected, the spin correlation functions near the two plaquettes ($W_{1}$ and $W_{2}$), are significantly suppressed, as shown in Fig. 5(b) and 5(c).
We also evaluate the energy expectation value of $|\Psi_2\rangle$ using the measured spin correlation functions,
\[{\langle{}E\rangle}_{\text{exp}}=-4.6494 \,(\pm0.1617) \,J.\] 
While the exact value is
\[{\langle{}E\rangle}_{\text{exact}}=-5.4641\,J.\]
The flux measurement results, presented in Fig. 5(d), confirm the successful annihilation of the vison pair.
Finally, the (quasi) probability distribution of the prepared state $|\Psi_2\rangle$ is shown in Fig. 5(e).

\subsection{Majorana fermion control}

In the final step, from the state $|\Psi_{2}\rangle$, we access the fermionic excited state, where the lowest fermion mode is annihilated and the fourth-lowest mode is created:
\[|1,0,0,0\rangle{}\mapsto{}|0,0,0,1\rangle{},\]
as illustrated in Fig. 6(a). 
Due to the degeneracy of the second and third lowest fermion modes in the two-vison sector, we excite the fourth-lowest mode to avoid ambiguity in reproducing the experiment.

We then repeat the same set of measurements to characterize the resulting state $|\Psi_3\rangle$. 
As in the vison manipulation process, we observe a strong suppression of spin correlation functions near the two plaquettes, as shown in Fig. 6(b) and 6(c). 
However, a clear distinction appears in the negative spin correlation functions, which reflect the increased energy and changed fermionic occupation of the state.
We evaluate the energy expectation value of $|\Psi_3\rangle$,
\[{\langle{}E\rangle}_{\text{exp}}=-1.47314 \,(\pm0.1710) \,J.\] 
While the exact value is
\[{\langle{}E\rangle}_{\text{exact}}=-1.4641\,J.\]
Next, the flux measurement shows that this process preserves vison configuration, as illustrated in Fig 6(d).
Finally, the (quasi) probability distribution of the state $|\Psi_3\rangle$ is shown in Fig. 6(e).

\begin{figure}[t]
\setcounter{figure}{6}
\includegraphics[width=1.0\linewidth]{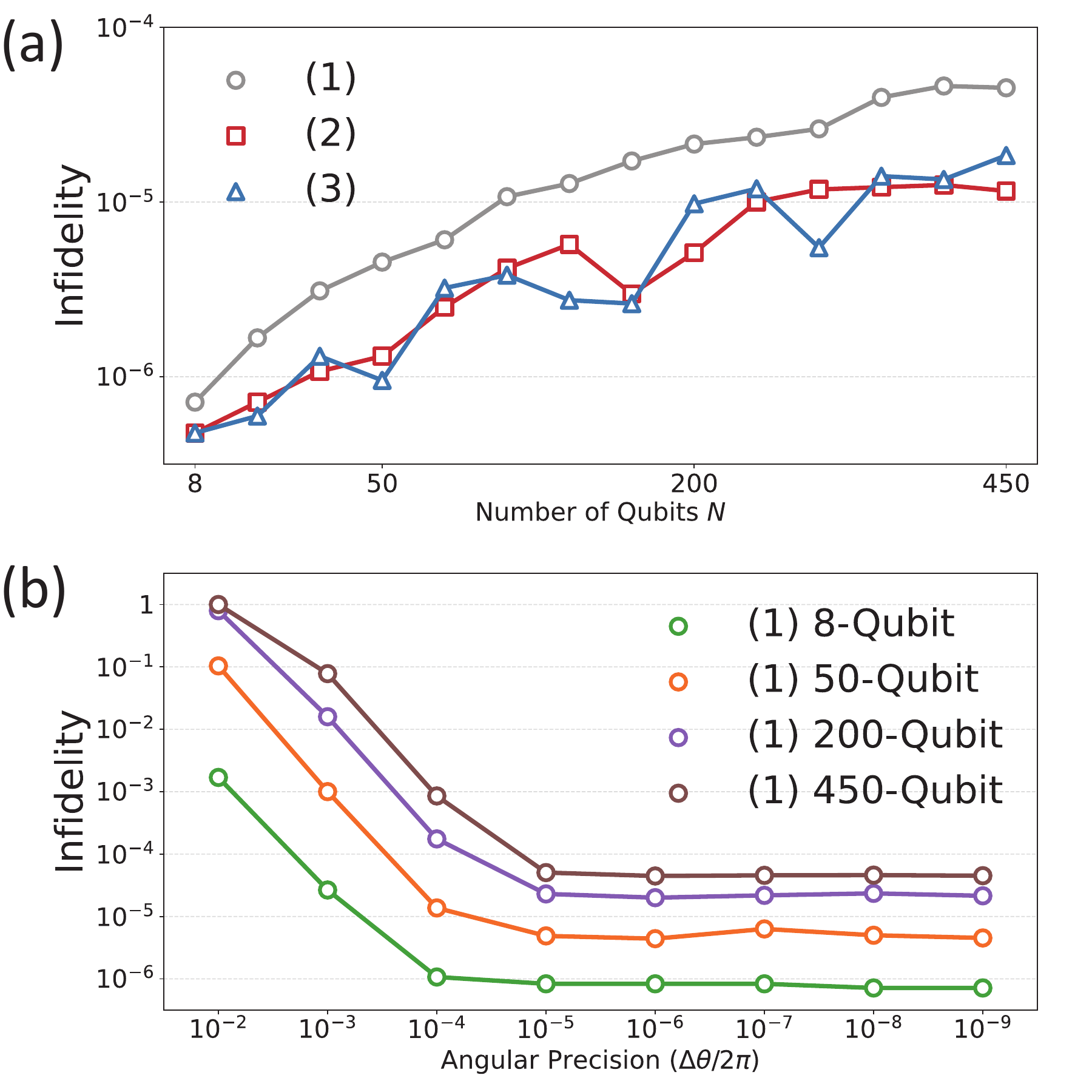}
\caption{Numerical simulation
(a) Numerically obtained infidelity between created and target states for each process as a function of system size. 
The length $L$ controls the system size ($L_1=L_2=L$ and $M=0$), from $L=2$ to $L=15$, the total number of spin is ($2 L^{2}$). 
First, in the process (1), the $U_{1}$ operator is designed to create KQSL GS in the vison-free sector. 
Second, in the process (2), the $U_{1}^{\prime}$ operator is designed to create an adjacent vison pair from the KQSL ground state. 
The corresponding $U_2$ operator is $\sigma_{i}^{z}$. 
Lastly, in the process (3), the $U_{1}^{\prime\prime}$ operator is designed to access the fermionic excited state from the fermionic ground state created from process (2).
For this figure, an angular precision on the order of $\Delta\theta/2\pi=10^{-9}$ was employed.
(b) Numerically obtained infidelity between created and target states for GS preparation as a function of angular precision $\Delta\theta$. 
}
\end{figure}

\section{Numerical Calculations}
Our proposal for large-scale KQSL quantum states can be further tested by performing numerical calculations, even though digital quantum simulation on a specific quantum platform is limited by its decoherence and system imperfection.  
Varying with the number of qubits, the three steps of our proposal have been realized. To be specific,  we consider the model with $L_1 = L_2 = L$ and $M = 0$ ($K/J = 0.1$). For varying $L$, the total number of spins is given by $2 L^2$.

We construct the quantum circuit for each process and measure the infidelity between the created and target quantum states.
This section provides the precise definition of the infidelity measure, along with a description of how it is computed.
The big advantage of this calculation is its scalability, which is obtained from fixing the $Z_2$ gauge field with a fixed vison configuration. 
Thus, one can expand this calculation with a much larger system. 
We constructed the unitary operator with varying system sizes for processes (1), (2), and (3), see Fig. 7(a). 
We confirm that our calculation gives the infidelity less than $10^{-4}$ with the system size up to 450 qubits.

Accurate realization of the target state requires high angular precision in the RZ gate operations, and this requirement becomes more significant as the system size increases, see Fig. 7(b).
In principle, our theoretical framework predicts that the infidelity should converge to zero as the angular precision increases. 
However, due to the limitations in numerical precision, infidelity exhibits a saturating behavior.

\subsection{Infidelity measure}
At each step, to implement the $U_{1}$ operator, we successively apply the $R$ operations.
\begin{gather*}
\mathbf{R}_{k}=\prod_{i=1}^{k} \otimes{}R(n_{i},\theta_{i}),\quad \mathbf{R}_{0}=I,\text{ and } \mathbf{R}_{M}=U_{1}.
\end{gather*}
We can check the performance at each step by measuring the infidelity between the created and target states. 
For instance, we have the initial state $|(\Phi_{\text{i}},f_{{\text{i}},\text{gs}})\rangle$ and the target state $U_{2}^{\dagger}|(\Phi_{\text{f}},f_{{\text{f}},\text{gs}})\rangle$ for the vision manipulation process. 
$U_2$ is identity if the process does not involve a change in vison configuration. 
\begin{equation}
C(k)=1-|\langle(\Phi_{\text{f}},f_{{\text{f}},\text{gs}})|U_{2}\mathbf{R}_{k}|(\Phi_{\text{i}},f_{{\text{i}},\text{gs}})\rangle|^{2}.
\label{Eq:infidelity}
\end{equation}
If the $U_{1}$ operator can be decomposed with $M$ operations, $C(M)$ becomes zero. 

We now rewrite this expression as an infidelity between different fermionic ground states in the same vison sector.
\begin{gather*}
    C(k) =1-|\langle(\Phi_{\text{i}},f_{{\text{i}},\text{gs}}^{\text{L}})|(\Phi_{\text{i}},f_{{\text{i}},\text{gs}}^{\text{R}})\rangle|^{2}.
\end{gather*}
Here, state $|\Phi_{\text{i}},f_{{\text{i}},\text{gs}}^{\text{L}}\rangle$ corresponds to the fermionic ground state of the quadratic Hamiltonian associated with the spin Hamiltonian $H_{\text{K}}^{\prime}=U_{2}^{\dagger} H_{\text{K}} U_{2}$ in the vison sector $\Phi_{\text{i}}$.
On the other hand, state $|\Phi_{\text{i}},f_{{\text{i}},\text{gs}}^{\text{R}}\rangle$ is identified as the fermionic ground state associated with the spin Hamiltonian $H_{\text{K}}$ in the vison sector $\Phi_{\text{i}}$, represented in the rotated frame.

\subsection{Overlap Calculation}

We show the explicit steps to calculate the overlap between two eigenstates ($|(\Phi_{i},f_{i}^{\text{L}})\rangle{}$ and $|(\Phi_{i},f_{i}^{\text{R}})\rangle{}$) of different quadratic Hamiltonian $H_{\text{L}}$ and $H_{\text{R}}$, provided that two states belong to the same vison sector $\Phi_{i}$. 
One can find detailed proof and related discussion in \cite{Knolle_2014,Bolukbasi_2012}.

This calculation aims to write two eigenstates in terms of the reference basis to obtain the overlap between the two states.
We apply the following transformation to $Q$. 
The matrix $Q_{\text{L}}$ ($Q_{\text{R}}$) is an orthogonal matrix obtained from the decomposition of quadratic Hamiltonian $H_{\text{L}}$ ($H_{\text{R}}$) \eqref{Eq:decomposition}.
\begin{equation}
    \begin{bmatrix}
        \overline{U}_{\text{L}}&V_{\text{L}}\\
        \overline{V}_{\text{L}}&U_{\text{L}}
    \end{bmatrix}=\frac{1}{2}\,\Gamma\, Q_{\text{L}}\, \Gamma^{\dagger}
    ,\quad\begin{bmatrix}
        \overline{U}_{\text{R}}&V_{\text{R}}\\
        \overline{V}_{\text{R}}&U_{\text{R}}
    \end{bmatrix}=\frac{1}{2}\,\Gamma\, Q_{\text{R}}\, \Gamma^{\dagger}.
\end{equation}
The matrix $\Gamma$ gives basis transformation from real fermion to complex fermion.
\begin{equation}
    \Gamma=
    \begin{bmatrix}
        +1&+i&0&0&0&0&\ldots\\
        0&0&+1&+i&0&0&\ldots\\
        0&0&0&0&+1&+i&\ldots\\
        \vdots&&&\ddots&&&\\
        +1&-i&0&0&0&0&\ldots\\
        0&0&+1&-i&0&0&\ldots\\
        0&0&0&0&+1&-i&\ldots\\
    \end{bmatrix}.
\end{equation}
On a particular reference basis ($U_{0}$ and $V_{0}$), they can be written as follows.
The two matrices, ${U}_{\text{L},0}$ and ${V}_{\text{L},0}$ (${U}_{\text{R},0}$ and ${V}_{\text{R},0}$), represent the canonical fermionic modes of the quadratic Hamiltonian $H_{\text{L}}$ ($H_{\text{R}}$), on a reference frame.
\begin{equation}
\begin{aligned}
{U}_{\text{L},0}=U_{0}^{\dagger}{U}_{\text{L}}+V_{0}^{\dagger}{V}_{\text{L}},\\
{V}_{\text{L},0}=V_{0}^{T}{U}_{\text{L}}+U_{0}^{T}{V}_{\text{L}},\\
{U}_{\text{R},0}=U_{0}^{\dagger}{U}_{\text{R}}+V_{0}^{\dagger}{V}_{\text{R}},\\
{V}_{\text{R},0}=V_{0}^{T}{U}_{\text{R}}+U_{0}^{T}{V}_{\text{R}}.
\end{aligned}
\end{equation}

If two states are the `fermionic vacuum' states of their respective quadratic Hamiltonians, their overlap can be calculated as follows.
\begin{equation}{|\langle(\Phi_{i},f_{i}^{\text{L}})|(\Phi_{i},f_{i}^{\text{R}})\rangle|}^{2}=|\text{det}({U}_{\text{R},0}^{\dagger}{U}_{\text{L},0}+{V}_{\text{R},0}^{\dagger}{V}_{\text{L},0})|.
\end{equation}
One can use this relation to calculate the overlap between fermionic excited states. 
Suppose that $|(\Phi_{i},f_{i}^{\text{R}})\rangle$ is the lowest energy fermion-occupied state instead of the vacuum state. 
Replacing the $Q_{\text{R}}$ with $Q_{\text{R}}T_{1,2}$ gives the correct overlap value for this case ($T_{1,2}$ is an elementary row operation that swaps first and second row). Physically, applying $T_{1,2}$ swaps the fermionic creation/annihilation operator ($a_{i,1}\leftrightarrow a_{i,1}^{\dagger}$). 
Thus, one can interpret the single fermion-occupied state as a fermionic vacuum state of other quadratic Hamiltonian.
One can further generalize this approach to obtain the overlap between two arbitrary fermionic excited states.

\section{Discussion and Conclusion}
In this work, we construct an exact unitary operator capable of preparing and controlling two quasiparticle excitations in the KQSL Hamiltonian at the level of a digital quantum simulator.
The construction of the unitary operator is based on two essential ideas.
First, we decompose the entire unitary operator into two separate components: the $U_{1}$ operator that gives rotation on fermion degree without changing vison configuration, and the $U_2$ (or $S$ operator for GS preparation) operator that changes vison configuration. 
This decomposition transforms the problem of connecting eigenstates in different vison sectors into the problem of connecting eigenstates within the same vison sector.
Second, we construct the $U_1$ operator in the fermionic representation and then translate it into a quantum circuit representation.
Since the $U_{1}$ operator does not change the vison sector, we restrict the unitary rotation within a specific subspace, effectively replacing an exponentially costly problem with one whose cost grows polynomially.
In exchange for obtaining scalability, the theory acts only within the exactly solvable limit.

We test our theory using the quantum processor to implement the designed quantum circuit. 
With the eight-qubit KQSL model, we successfully demonstrate the preparation of the KQSL ground state and independent control of two quasiparticle excitations.
We verified the properties of the prepared quantum state through the vison measurement, spin correlation function analysis, and quantum state tomography.

We expect our results to serve as a guiding protocol for future digital quantum simulations of the KQSL Hamiltonian on other hardware platforms.
Our protocol is tested for larger number of qubits (12 and 18 qubits), but we could only obtain meaningful experimental data in GS preparation for the 12-qubit model, see Appendix C.
To elaborate, while our numerical simulation results (up to 450 qubits) demonstrate the scalability of our protocol in an ideal setting, the experimental realization with current NISQ hardware shows practical limitations.
The primary source of this inconsistency is the accumulated gate errors.
Although the circuit depth to implement the $U_1$ operator scales as $O(N)$, the total number of two-qubit gates scales as $O(N^2)$.
Given that the current hardware device has finite gate fidelity, the overall fidelity of the prepared state decreases significantly with increasing system size.
Additionally, we employ a multi-qubit Pauli measurement to characterize the KQSL eigenstate, which is particularly sensitive to errors.
The vison configuration is characterized by plaquette operators ($W_p$), which are six-qubit Pauli strings.
For instance, a single Pauli error on the plaquette can flip the measured vison value, leading to an incorrect identification of the topological sector.
A more detailed analysis of the error channels and their impact is provided in Appendix E.

This vulnerability to errors poses a significant challenge to the practical scalability of our protocol.
For future work, various error mitigation (and correction) can be applied to suppress the impact of such errors. 
It is worth referring to the error handling strategies used in similar experimental studies \cite{Evered_2025,Will_2025}, which first prepared a vison-free state and employed unitary evolution conserving vison configuration.
These studies employed multiple methods, including measurement-based state preparation, mid-circuit feedforward, post-selection of data, and randomized compiling.
Depending on the properties of the hardware, one should select the proper method, and it will be necessary to employ them to realize the KQSL eigenstates in larger systems.

\section{Acknowledgments}
This work was supported by the National Research Foundation of Korea (NRF) grant funded by the Korean government [Ministry of Science and ICT (MSIT)] (Grant Nos. RS-2025-00559286, 2022M3H4A1A04074153), the Nano \& Material Technology Development Program through the National Research Foundation of Korea (NRF) funded by Ministry of Science and ICT (MSIT) (Grant Nos. RS-2023-00281839, RS-2024-00451261), National Measurement Standards Services and Technical Support for Industries funded by Korea Research Institute of Standards and Science (KRISS) (Grant No. KRISS–2025–GP2025-0015), ‘Quantum Information Science R\&D Ecosystem Creation’ through the National Research Foundation of Korea (NRF) funded by the Korean government [Ministry of Science and ICT (MSIT)] (Grant No. 2020M3H3A1110365).


\appendix
\section{Exact solution of the KQSL Hamiltonian}  
\subsection{Exact solution of the Hamiltonian}
\begin{gather*}
    H_{\text{K}}=-\sum_{\langle{}i,j\rangle}J_{ij}^{\alpha}\sigma_{i}^{\alpha}\sigma_{j}^{\alpha}-\sum_{(i,j,k)}K_{ijk}\sigma_{i}^{x}\sigma_{j}^{y}\sigma_{k}^{z}.
\end{gather*}

To solve the Hamiltonian, we map the spin Hamiltonian to a quadratic form by applying the Kitaev transformation, $\sigma_i^\alpha \to i b_i^\alpha c_i$, where $b_i^x$, $b_i^y$, $b_i^z$, and $c_i$ are Majorana fermion operators.
\begin{equation}
H(A)=\frac{i}{4}\sum_{i,j}A_{ij}c_{i}c_{j},\ A_{ij}=2J_{ij}u_{ij}+\sum_{k}2K_{ijk}u_{ik}u_{jk},
\label{Eq:H_Quadartic}
\end{equation}
where $u_{ij}=i{b}_{i}^{\alpha}{b}_{j}^{\alpha}$. 
One can use the eigenvalue of $u_{ij}$ to divide the total Hilbert space into a set of subspaces. 
Here, $u_{ij}$ act as $Z_2$ gauge field.
Thus, itself is not a gauge invariant operator: $u_{ij}$ is not a physical observable. 
Instead, a product of $u_{ij}s$ along a closed path is gauge invariant. 
One can define a fermionic path operator as follows. 
\begin{equation}
W(i_{1},i_{2},\dotsc,i_{n})=\sigma_{i_{n}}^{\alpha_{n-1,n}}\sigma_{i_{n-1}}^{\alpha_{n-1,n}}\dotsc\sigma_{i_{2}}^{\alpha_{1,2}}\sigma_{i_{1}}^{\alpha_{1,2}}.
\label{Eq:Path_operator}
\end{equation}
$\{i_{1},i_{2},\dotsc,i_{n}\}$ is ordered path defined on honeycomb lattice. 
The fermionic path operator defined on the closed path commutes with Hamiltonian, and it can be written as the product of $u_{ij}s$ along the path. 
For example, one can define the $Z_2$ flux operator on a single plaquette as follows; for the explicit index notation, see Fig. 8(a).
\begin{gather*}
\begin{aligned}
W_{p}&=\sigma_{1}^{z}\sigma_{2}^{x}\sigma_{3}^{y}\sigma_{4}^{z}\sigma_{5}^{x}\sigma_{6}^{y}\\
&=-u_{12}u_{23}u_{34}u_{45}u_{56}u_{61}.
\end{aligned}
\end{gather*}
 
One can easily diagonalize the Hamiltonian within the subspace characterized by fixed $u_{ij}$ configuration.
\begin{gather*}
({b}_{1}^{\prime},{b}_{1}^{\prime\prime},\dotsc,{b}_{N/2}^{\prime},{b}_{N/2}^{\prime\prime})=({c}_{1},{c}_{2},\dotsc,{c}_{N-1},{c}_{N})Q,
\label{Eq:Q_definition}
\end{gather*}
where $Q$ satisfies,
\begin{equation}
    \label{Eq:decomposition}
A=Q\begin{bmatrix}
    0 & +\epsilon_{1} &  & &\\
    -\epsilon_{1} & 0& &  & \\
    & & \ddots&  & \\
    & & & 0 & +\epsilon_{N/2} \\
    & & & -\epsilon_{N/2} & 0 
   \end{bmatrix}Q^{T}.
\end{equation}
$\pm\epsilon_{k}$ is an eigenvalue of $iA$, odd (even) columns of $Q$ are real (imaginary) parts of the eigenvectors. 
The $N$ is the number of qubits (spins), which is an even number.
All $\epsilon_{k}$s are non-negative and ordered in increasing order ($0\leq\epsilon_{1}\leq\epsilon_{2}\leq\dotsc\leq\epsilon_{N/2}$). 
The canonical form of quadratic Hamiltonian is
\begin{equation}
H(A)=\frac{i}{2}\sum_{k=1}^{N/2}\epsilon_{k}{b}_{k}^{\prime}{b}_{k}^{\prime\prime}=\sum_{k=1}^{N/2}\epsilon_{k}(a_{k}^{^{\dagger}}a_{k}-\frac{1}{2}).
\label{Eq:H_canonical}
\end{equation}
$a_{k}^{\dagger}=\frac{1}{2}({b}_{k}^{\prime}-i{b}_{k}^{\prime\prime})$ and $a_{k}=\frac{1}{2}({b}_{k}^{\prime}+i{b}_{k}^{\prime\prime})$. 
Thus, every eigenstate $|(\Phi,f)\rangle$ is labeled by two excitations: $\Phi$, set by the vison configuration, and $f$, describing fermionic excitations of the quadratic Hamiltonian.

The matrix $Q$ encodes all necessary information about the fermionic excitations of quadratic Hamiltonian $H(A)$; it is a Bogoliubov matrix written on a Majorana fermion basis and will play a central role in the following discussion.

\subsection{Projection operator \& physical fermion parity}

\begin{figure*}[t]
\setcounter{figure}{7}
\includegraphics[width=1.0\linewidth]{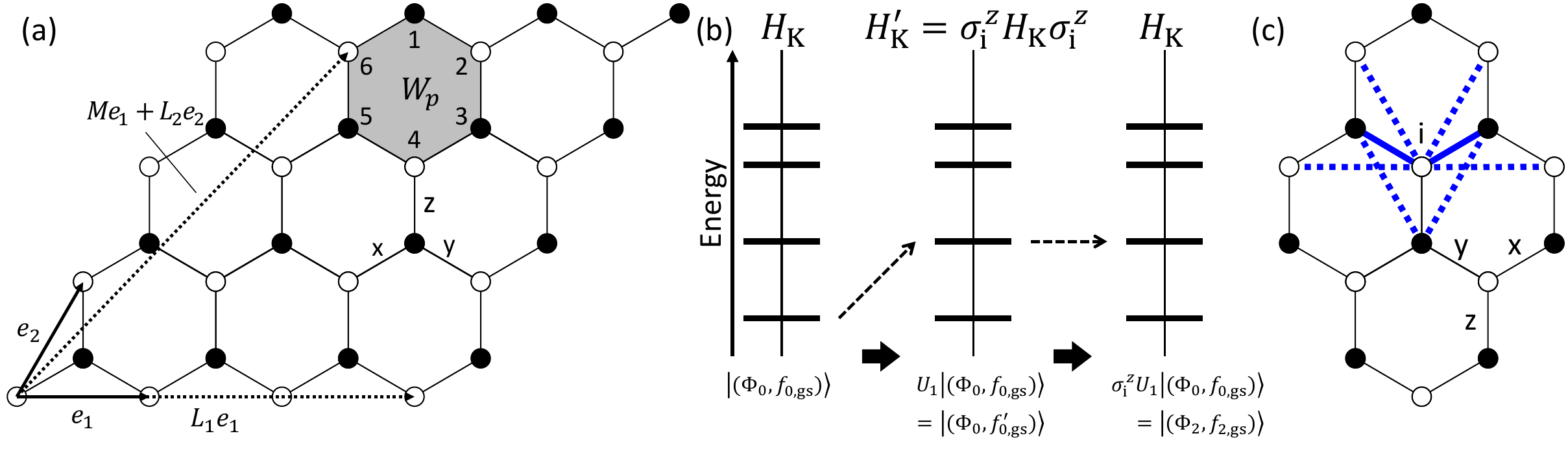}
\caption{(a) The geometry of Kitaev honeycomb model on the torus. 
Dashed arrows connect the identical sites on the torus, ($L_1=L_2=3$ and $M=1$). $M$ is twisting parameter of torus. 
(b) The construction of unitary operator connecting the eigenstates in different vison sectors. 
(c) Applying $U_{2}=\sigma_{i}^z$ on both side of $H_{\text{K}}$ flips two $J$ links and six $K$ links in Hamiltonian.}
\end{figure*}
As the transformation from spin to Majorana fermion doubles the dimension of Hilbert space, the projection operator onto the physical Hilbert space is required to remove the unphysical eigenstates. 
The projection operator onto physical Hilbert space is
\begin{gather*}
    P=\prod_{j}\frac{1+D_{j}}{2}, \quad D_{j}= {b_{j}}^{x}{b_{j}}^{y}{b_{j}}^{z}c_{j}.
\end{gather*}

The exact evaluation of the projection operator was first done in Ref. \cite{Loss_2011}, with the following decomposition. 
\begin{equation}
    P=\left(\cfrac{1}{2^{N-1}}\sum_{\{j\}}\prod_{j\in{}\{j\}}D_{j}\right)\left(\cfrac{1+\prod_{i}^{N}D_{i}}{2}\right)=S\cdot{}P_{0}.
    \label{Eq:projector_0}
\end{equation}

The first part ($S$) symmetrizes the state, connecting all gauge equivalent states ($\{j\}$ indicates all subsets of index set, if $\{j\}$ is included, then $D-\{j\}$ is not). 
The second part, $P_{0}$, determines whether the state is physical or not. Then, $P_{0}$ can be written in the following way.
\begin{equation}
    2P_{0}=1+{(-1)}^{L_{1}+L_{2}+M(L_{1}-M)}\hat{\pi}_{\text{phy}}\text{det}(Q)\prod_{\langle{}i,j\rangle}u_{ij}.
    \label{Eq:Projector_phy}
\end{equation}
$L_1$ and $L_2$ are the lengths of the system ($L_1\times L_2$ unit cells in total), $M$ is the twisting parameter of a torus (see Fig. 8(a)), $Q$ is the orthogonal matrix obtained from \eqref{Eq:decomposition}, and $\hat{\pi}_{\text{phy}}$ is the physical fermion parity operator.
\begin{equation}
    \hat{\pi}_{\text{phy}}=\text{det}(Q)\big[(-i)^{N/2}\prod_{i}c_{i}\big]=\text{det}(Q) \hat{\pi}_{c}.
    \label{Eq:fermion_parity_operator}
\end{equation}

As a result, only the even (or odd) modes become physical states depending on the vison configuration and geometry. 
The physical fermion parity $\pi_{\text{phy}}$ is gauge invariant quantity. 
Note that $\text{det}(Q)$ and $\prod_{}u_{ij}$ are not gauge invariant, but $(\text{det}(Q)\prod_{}u_{ij})$ is a gauge invariant quantity.

The concept of physical fermion parity removes the possible ambiguities in counting fermion excitations. 
Thus, it is an essential concept to describe the fermionic excitations in gauge invariant language. 
For example, consider the state where the fermion mode is occupied with the corresponding mode energy $(+\epsilon)$. 
Alternatively, one can interpret this state as an unoccupied fermion state with the corresponding mode energy $(-\epsilon)$.
As the Eq. \eqref{Eq:decomposition} imposes the condition that every fermion mode has a positive energy, physical fermion parity always counts the number of positive energy fermions, regardless of gauge choice.

One of the necessary conditions for constructing the unitary operator connecting the eigenstates of KQSL is that the physical fermion parity must be determined for the initial and final states.
This is always possible as long as we do not have any gapless energy mode ($\epsilon_{i}>0$ for all $i$). 
If the gapless mode exists, additional procedures are required to determine the physical fermion parity.
In order to open the gap, one can add a small perturbation on the parameters $J$ and $K$, either locally or globally, then take the zero perturbation limit to recover the original Hamiltonian without the ambiguity in the physical fermion parity.

\section{Construction and Implementation of Unitary Operator}
In this study, we will explain how to construct the exact unitary operator with four specific examples. 
The first example is (i) ground state preparation that maps the initial state to the ground state $|(\Phi_{\text{0}},f_{{\text{0}},\text{gs}})\rangle$ of KQSL, provided that the initial state lies in the vison sector $\Phi_{\text{0}}$.
Next is the (ii) vison manipulation, the process that maps $(|\Phi_{\text{0}},f_{{\text{0}},\text{gs}})\rangle$ to $(|\Phi_{\text{2}},f_{{\text{2}},\text{gs}})\rangle$ (creation of adjacent vison pair).
This example can be further generalized to the mapping between arbitrary vison configurations.
The third example is (iii) Majorana fermion control, which allows us to access the fermionic excited states without changing the vison configuration. 
The last example is (iv) Majorana fermion readout, which gives the information of the fermion occupation number.

Then, we will explicitly show how the $U_{1}$ operator, which involves nonlocal many-qubit operations, can be decomposed into a set of local unitary gate operations.

\subsection{Algebraic properties of unitary operator}
Again, we asserts that the following unitary operator can connect the two arbitrary eigenstates of KQSL.
\begin{gather*}
|(\Phi_{\text{f}},f_{{\text{f}}})\rangle=U_{2}U_{1}|(\Phi_{\text{i}},f_{{\text{i}}})\rangle
\end{gather*}
In this section, we will discuss some important properties of the unitary operator before dealing with specific examples.

\subsubsection{$U_{1}$ operator: Rotation in the fermion space}
As the $U_1$ operator \eqref{Eq:U_1_B} does not change the vison configuration, it is safe to fix the $Z_2$ gauge field with the vison sector $\Phi_{\text{i}}$. 
Suppose we have the quadratic Hamiltonian $H(A_{\text{i}})$ ($A_{\text{i}}=Q_{\text{i}}E_{\text{i}}Q_{\text{i}}^{T}$) obtained from assigning specific $Z_2$ gauge field for vison sector of initial state $|(\Phi_{\text{i}},f_{{\text{i}}})\rangle$. 
Then, we consider the unitary transformation applied to quadratic Hamiltonian, $H(A_{\text{i}})\to{}e^{-iH(B)}[H(A_{\text{i}})]e^{+iH(B)}$. 
In the definition of quadratic Hamiltonian \eqref{Eq:H_Quadartic}, Kitaev~\cite{Kitaev_2006} used a factor of $1/4$ to have the following commutation relation between quadratic Hamiltonians,
\begin{gather*}
   [-iH(A),-iH(B)]=-iH([A,B]).
\end{gather*}
This relation is useful for obtaining the transformed quadratic Hamiltonian.
\begin{gather*}
    \begin{aligned}
    &e^{-iH(B)}[-iH(A)]e^{+iH(B)}\\
    =&-iH(A+[B,A]+\frac{1}{2}[B,[B,A]]+\dotsc)\\
    =&-iH(e^{+B}Ae^{-B}).
    \end{aligned}
\end{gather*}
As a result, the quadratic Hamiltonian $H(A_{\text{i}})$ transforms to $H(A_{\text{f}})$ ($A_{\text{f}}=(e^{+B}Q_{\text{i}})E_{\text{i}}(e^{+B}Q_{\text{i}})^{T}$). 
$e^{+B}$ is a special orthogonal matrix that characterizes the unitary rotation. 
Physically, the $U_{1}$ operator can be understood as a time evolution operator acting on fermion space, with the quadratic Hamiltonian $H(B)$.

In this study, we first create the $U_{1}$ operator on the fermionic basis, then reconstruct it on the qubit (spin) basis.
As the transformation from spin to Majorana fermion doubles the dimension of Hilbert space, the projection operator onto the physical Hilbert space is required to remove the unphysical eigenstates. 
The exact evaluation of the projection operator was first done in \cite{Loss_2011}. 
They introduced the concept of physical fermion parity to describe the physical eigenstates in a gauge invariant language. 
As the physical fermion parity removes possible ambiguities in identifying physical eigenstates, it plays a crucial role in constructing the unitary operator; to construct the unitary operator, the physical fermion parity must be determined for the initial and final state.

\subsubsection{$U_{2}$ operator: vison manipulation}

$U_2$ operator is a Pauli string operator connecting the vison sectors $\Phi_{\text{i}}$ and $\Phi_{\text{f}}$. 
As we discussed, the Majorana fermion is defined on $Z_{2}$ gauge field characterized by vison configuration. 
Thus, the local unitary operator that affects vison degree of freedom also affects the Majorana fermion indirectly by changing $Z_2$ gauge field. 
For instance, consider introducing ${\sigma}_{i}^{z}$ to the fermionic ground state on the vison-free sector $|(\Phi_{\text{0}},f_{{\text{0}},\text{gs}})\rangle$: fermionic vacuum or the lowest fermion mode is occupied depending on allowed physical fermion parity. 
It will create an adjacent vison pair and simultaneously rearrange the Majorana fermion as the gauge field changes from a vison-free sector to a two-vison sector \cite{Baskaran_2007}. 
As the local spin operator affects the vison and fermion degree of freedom,  approximating $|(\Phi_{\text{2}},f_{{\text{2}},\text{gs}})\rangle$ by  ${\sigma}_{i}^{z}|(\Phi_{\text{0}},f_{{\text{0}},\text{gs}})\rangle$ gives poor fidelity.

The main trick to tackle this problem is to apply the $U_{2}$ operator after applying the $U_{1}$ operator, making the $U_{2}^{\dagger}|(\Phi_{\text{f}},f_{{\text{f}}})\rangle$ belongs to the initial vison sector. 
While $|(\Phi_{\text{f}},f_{{\text{f}}})\rangle$ is the eigenstate of $H_{\text{K}}$ (the original KQSL Hamiltonian) with the vison configuration $\Phi_{\text{f}}$, $U_{2}^{\dagger}|(\Phi_{\text{f}},f_{{\text{f}}})\rangle=|(\Phi_{\text{i}},f_{{\text{i}}}^{\prime})\rangle$ is the eigenstate of $H_{\text{K}}^{\prime}=U_{2}^{\dagger} H_{\text{K}} U_{2}$ with the vison configuration $\Phi_{\text{i}}$.
As the two states  $|(\Phi_{\text{i}},f_{{\text{i}}})\rangle$ and $U_{2}^{\dagger}|(\Phi_{\text{f}},f_{{\text{f}}})\rangle$ belong to the same vison sector, it is possible to connect two states with the $U_1$ operator. 
The Hamiltonian $H_{\text{K}}^{\prime}$ remains exactly solvable. For example, if $U_{2}=\sigma_{i}^z$, the sign of two $J$ and six $K$ links will be flipped as in Fig. 8(c). 
Through this strategy, we transform the problem of connecting the eigenstates in different vison sectors into a problem of connecting eigenstates in the same vison sector, see Fig. 8(b).
\begin{gather*}
    |(\Phi_{\text{i}},f_{{\text{i}},\text{gs}})\rangle \mapsto
U_{2}^{\dagger}|(\Phi_{\text{f}},f_{{\text{f}},\text{gs}})\rangle=|(\Phi_{\text{i}},f_{{\text{i}},\text{gs}}^{\prime})\rangle.
\end{gather*}

\subsection{Ground state preparation}
We will show the explicit steps to construct the unitary operator that can be used in the ground state preparation process.
We first need to define the initial state. 
Starting from the product state ${|+x\rangle}^{N}$, we perform a projection operator to map the state to the given vison sector where the KQSL ground state $|(\Phi_{\text{0}},f_{{\text{0}},\text{gs}})\rangle$ lies.
Generally, the KQSL ground state lies on the vison-free sector, $w_{p}=+1$ for all plaquette.
\begin{equation}
|(\Phi_{\text{0}},f_{0,\text{gs}}^{\text{dim}})\rangle\propto{}\prod_{p}\frac{1+W_{p}}{2}\prod_{i=X,Y}\frac{1+{(-1)}^{w_{i}}W_{i}}{2}{|+x\rangle}^{N}.
\label{Eq:projection_to_GS}
\end{equation}
$W_{p}$ operators are plaquette operators, and $W_{X,Y}$ are fermionic path operators along two non-contractible loops of torus (known as Wilson loop variable). 
Operators in $\{W_{p}\}\cup\{W_{X},W_{Y}\}$ are mutually commuting; thus, the order in the product does not matter. 
As we have information of the initial state, these projection operators can be replaced with unitary gate operations (the $S$ operator).
\begin{gather*}
|(\Phi_{\text{0}},f_{0,\text{gs}}^{\text{dim}})\rangle=S{|0\rangle}^{N}.
\end{gather*}
This method is used to prepare Toric code ground state \cite{Satzinger_2021} and vison-free state \cite{Jin_2023, Will_2025} on the honeycomb lattice.
Depending on the model's geometry, the projection operator with $W_{X, Y}$ can be dropped.
While the torus model requires $W_{X}$ and $W_{Y}$, the strip model requires only one of them. 
\begin{equation}
\label{Eq:H_dim}
    H_{\text{dim}}=-\sum_{\langle{}i,j\rangle\in{}X}J_{ij}\sigma_{i}^{x}\sigma_{j}^{x}\quad{} (J_{ij}>0).
\end{equation}
The state $|(\Phi_{\text{0}},f_{0,\text{gs}}^{\text{dim}})\rangle$ can be considered as the eigenstate of the following Hamiltonian, with the vison configuration $\Phi_{\text{0}}$. 
The projection operator selects a certain ground state of $H_{\text{dim}}$ that lies on the same vison sector of the KQSL ground state.
If the allowed physical fermion parity is odd for vison sector $\Phi_{0}$, this projection operator gives simply 0. 
In the torus model, depending on the Wilson loop variable (and spatial periodicity $L_1$, $L_2$, and $M$), allowed physical fermion parity can be odd for the vison-free sector \cite{Vojta_2015}. 
In this case, one should start with the product state ${|-x\rangle}\otimes{|+x\rangle}^{N-1}$.
To avoid the degeneracy, one can use the site-dependent $J_{ij}$.

Now, we have two states, $|(\Phi_{0},f_{0,\text{gs}}^{\text{dim}})\rangle$  and $|(\Phi_{0},f_{{\text{0}},\text{gs}})\rangle$, on the same vison sector, their respective Hamiltonians ($H_{\text{dim}}$ and $H_{\text{K}}$), and the physical fermion parity determined for each state. 
We first choose a certain $Z_2$ gauge field to obtain the quadratic Hamiltonians: $H(A_{\text{dim}})$ and $H(A_{\text{0}})$ ($A_{\text{dim}}=Q_{\text{dim}}E_{\text{dim}}Q_{\text{dim}}^{T}$ and $A_{\text{0}}=Q_{\text{0}}E_{\text{0}}Q_{\text{0}}^{T}$).
Now, the construction rule changes depending on whether the two Hamiltonians allow the identical physical fermion parity (on the vison sector $\Phi_{\text{0}}$) or not. 
Note that ($f_{{\text{0}},\text{gs}}$) does not mean the fermionic vacuum; it can be either a fermionic vacuum or the lowest energy mode is occupied depending on allowed physical fermion parity.

First, we consider the case of two Hamiltonians allowing identical physical fermion parity.'
From the equation \eqref{Eq:Projector_phy}, it will naturally impose the condition $\text{det}(Q_{\text{dim}})=\text{det}(Q_{\text{0}})$. 
Note that $\text{det}(Q_{\text{dim}})$ and $\text{det}(Q_{\text{0}})$ are gauge-dependent quantities, but their relation is preserved under $Z_2$ gauge transformation. Then, we can determine the matrix $B$ as follows,
\begin{equation}
e^{B}=Q_{\text{0}}Q_{\text{dim}}^{T}\quad \text{(identical parity)}.
\label{Eq:B_identical_parity_GS}
\end{equation}
One can calculate the matrix $B$ by calculating the principal logarithm of $Q_{\text{0}}Q_{\text{dim}}^{T}$. 
Under the action of the $U_{1}$ operator, $H(A_{\text{dim}})$ transforms into $H(A_{\text{dim}}^{\prime})$ ($A_{\text{dim}}^{\prime}=Q_{\text{0}}E_{\text{dim}}{Q_{\text{0}}^{T}}$). As a result, the state $|(\Phi_{0},f_{0,\text{gs}}^{\text{dim}})\rangle$ is transformed to $|(\Phi_{0},f_{{\text{0}},\text{gs}})\rangle$. 
The $U_{1}$ operator gives the following rotation in fermion space,
\begin{equation}
\label{Eq:fermion_mode_rotation_equal}
(a_{\text{dim},1},a_{\text{dim},2},\dotsc,a_{\text{dim},N/2})\to(a_{\text{0},1},a_{\text{0},2},\dotsc,a_{\text{0},N/2}).
\end{equation}

The $U_1$ operator also ensures the mapping between excited states; if the initial state has the first and second lowest energy fermions of $H(A_{\text{dim}})$, the final state has the first and second lowest energy fermions of $H(A_{\text{0}})$.

Second, we need a modification if the initial and final vison sectors allow different physical fermion parity.
The Eq. \eqref{Eq:Projector_phy} impose the condition $\text{det}(Q_{\text{dim}})=-\text{det}(Q_{\text{0}})$. 
As the $Q_{\text{0}}Q_{\text{dim}}^{T}\notin{}SO(N)$, Eq. \eqref{Eq:B_identical_parity_GS} breaks down. 
In this case, we need a mapping that maps every even parity mode to an odd parity mode and vice versa. 
One can resolve this issue by adding the $T_{1,2}$ (elementary row operation that swaps first and second row) between $Q_{\text{0}}$ and $Q_{\text{dim}}^{T}$.
\begin{equation}
e^{B}=Q_{\text{0}}T_{1,2}Q_{\text{dim}}^{T}\quad \text{(different parity)}.
\label{Eq:B_different_parity_GS}
\end{equation}
We replace $Q_{\text{0}}$ by $Q_{\text{0}}T_{1,2}$. 
In other words, we swaps $a_{\text{0},1}$ and $a_{\text{0},1}^{\dagger}$. 
Unlike the previous case, the $U_{1}$ operator gives the following rotation in fermion space,
\begin{equation}
\label{Eq:fermion_mode_rotation_dif}
(a_{\text{dim},1},a_{\text{dim},2},\dotsc,a_{\text{dim},N/2})\to(a_{\text{0},1}^{\dagger},a_{\text{0},2},\dotsc,a_{\text{0},N/2}).
\end{equation}
Suppose the $H(A_{\text{dim}})$ allows even physical fermion modes, and $H(A_{\text{0}})$ allows odd physical fermion modes. 
Then, the $U_1$ operator will map the fermionic vacuum state of $H(A_{\text{dim}})$ to the lowest fermion mode occupied state of $H(A_{\text{0}})$. 
Again, the $U_1$ operator also ensures the mapping between excited states, but in a different manner. 
For instance,  if the initial state has the first and second lowest energy fermions of $H(A_{\text{dim}})$, the final state has a second lowest fermion of $H(A_{\text{0}})$.

One may have difficulty in determining physical fermion parity in the presence of gapless mode. 
A good example is the vison-free sector of the $B$ phase KQSL model (ex) $J=1$ and $K=0$). 
One can treat this problem by adding a small $K=\delta{}k$ and taking the $\delta{}k\rightarrow{}0$ limit to obtain the $Q$ matrix and physical fermion parity. 

You may wonder how the physical fermion parity can change while the explicit form of $U_{1}$ operator \eqref{Eq:U_1_B} imposes parity conservation. 
One should understand the subtle differences between physical fermion parity $\pi_{\text{phy}}$ and $c$ fermion parity $\pi_{\text{c}}$ \eqref{Eq:fermion_parity_operator}. $c$ fermion parity is a preserved quantity; one can check it from the commutator relation $[\hat{\pi}_{c},U_{1}]=0$. 
However, it is not a gauge invariant quantity. 
Thus, we use the physical fermion parity to describe the eigenstate in a gauge invariant language. 
The physical fermion parity became varying quantity in exchange for obtaining gauge invariance. 
In other words, the change in physical fermion parity may occur as the energy of certain fermion modes changes sign, without a change in $c$ fermion parity: what changes is not the fermion parity, but how to count fermion parity.

\begin{figure}[t]
\setcounter{figure}{8}
\includegraphics[width=1.0\linewidth]{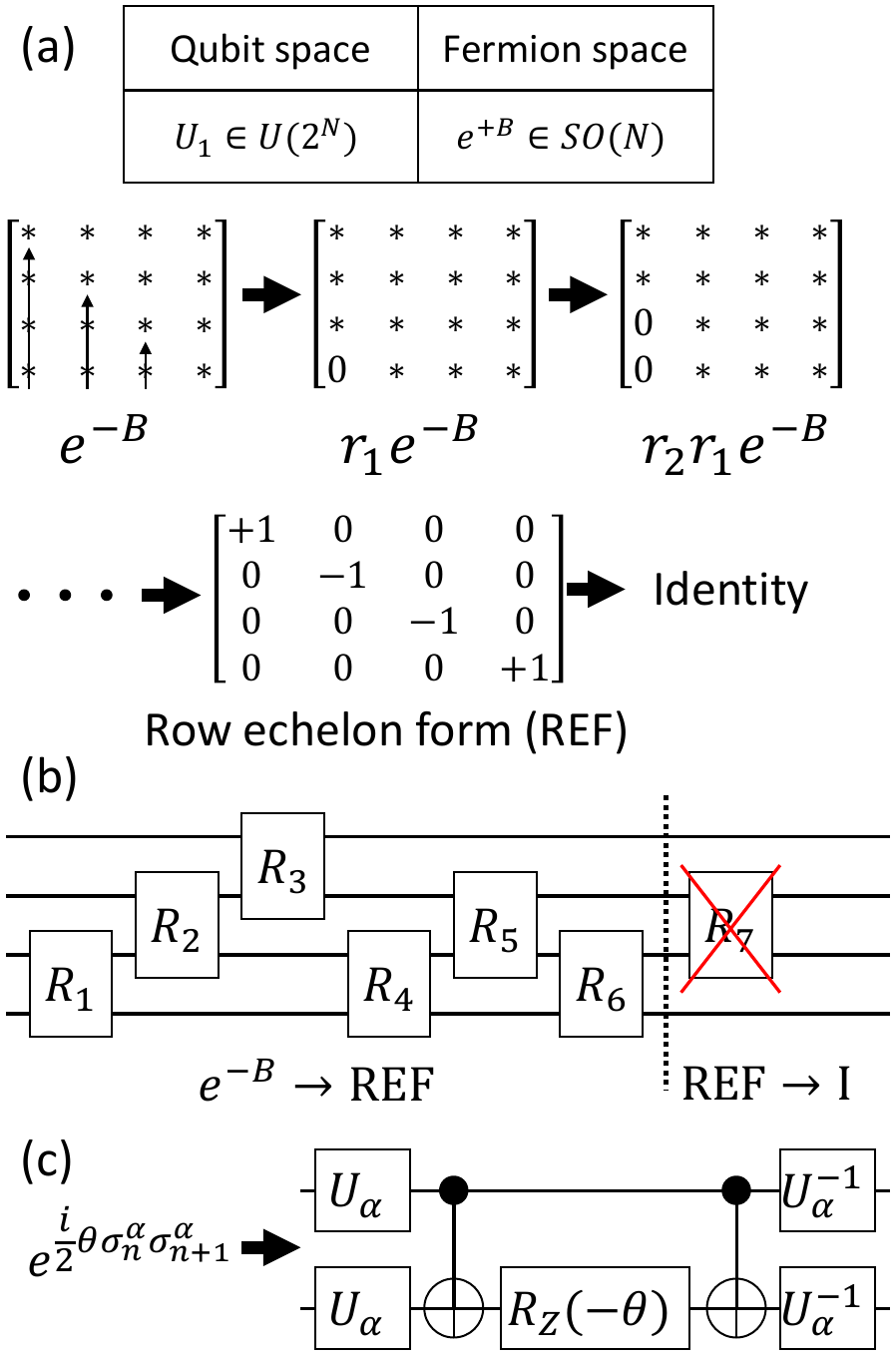}
\caption{(a) Explicit procedure for decomposing the matrix $e^{+B}$. 
(b) Quantum circuit representation of $U_{1}$ operator decomposition. 
The $R_{7}$ operation can be absorbed into $R_{5}$ operation.
(c) Implementation of $R$ operations.
$U_{x}=H$, $U_{y}=H S^{\dagger}$, and  $U_{z}=I$.
}
\end{figure}
\subsection{Vison manipulation}
We will show the explicit protocol to construct the unitary operator connecting two eigenstates, $|(\Phi_{\text{0}},f_{{\text{0}},\text{gs}})\rangle$ and $|(\Phi_{\text{2}},f_{{\text{2}},\text{gs}})\rangle$, the process to create the adjacent vison pair from the vison-free sector. 
Applying $U_{2}^{\dagger}$ to the final state makes the state $U_{2}^{\dagger}|(\Phi_{\text{f}},f_{{\text{f}},\text{gs}})\rangle$ belong to the initial vison sector, which allows us to find the $U_{1}$ operator connecting two states with a fixed $Z_2$ gauge field.

The initial state $|(\Phi_{\text{0}},f_{{\text{0}},\text{gs}})\rangle$ is an eigenstate of $H_{\text{K}}$ (KQSL Hamitonian) with vison configuration $\Phi_{\text{0}}$, and the final state $|(\Phi_{\text{2}},f_{{\text{2}},\text{gs}})\rangle$ is an eigenstate of $H_{\text{K}}$ with vison configuration $\Phi_{\text{2}}$. 
In this case, we can simply choose $U_2$ as a single Pauli matrix, $U_{2}=\sigma_{i}^{z}$. Then one can view $\sigma_{i}^{z}|(\Phi_{\text{2}},f_{{\text{2}},\text{gs}})\rangle$ as an eigenstate of $H_{\text{K}}^{\prime}=\sigma_{i}^{z}H_{\text{K}}\sigma_{i}^{z}$ with the vison configuration $\Phi_{\text{0}}$, $\sigma_{i}^{z}|(\Phi_{\text{2}},f_{{\text{2}},\text{gs}})\rangle=|(\Phi_{\text{0}},f_{{\text{0}},\text{gs}}^{\prime})\rangle$.
Treating $\sigma_{i}^{z}|(\Phi_{\text{2}},f_{{\text{2}},\text{gs}})\rangle$ as $|(\Phi_{\text{0}},f_{{\text{0}},\text{gs}}^{\prime})\rangle$ does not change the physical fermion parity. 
Two states, $|(\Phi_{\text{2}},f_{{\text{2}},\text{gs}})\rangle$ and $|(\Phi_{\text{0}},f_{{\text{0}},\text{gs}}^{\prime})\rangle$ have identical energy and physical fermion parity with their respective Hamiltonians ($H_{\text{K}}$ and $H_{\text{K}}^{\prime}$) on respective vison sectors ($\Phi_{\text{2}}$ and $\Phi_{\text{0}}$).

Now, we have two states, $|(\Phi_{\text{0}},f_{{\text{0}},\text{gs}})\rangle$ and $|(\Phi_{\text{0}},f_{{\text{0}},\text{gs}}^{\prime})\rangle$, on the same vison sector, their respective Hamiltonians ($H_{\text{K}}$ and $H_{\text{K}}^{\prime}$, and the physical fermion parity determined for each state. 
We first choose a certain $Z_2$ gauge field to obtain the quadratic Hamiltonians: $H(A_{\text{0}})$ and $H(A_{\text{0}}^{\prime})$ ($A_{\text{0}}=Q_{\text{0}}E_{\text{0}}Q_{\text{0}}^{T}$ and $A_{\text{0}}^{\prime}=Q_{\text{0}}^{\prime}E_{\text{0}}^{\prime}{Q_{\text{0}}^{\prime T}}$).
We transform the problem of connecting two states ($|(\Phi_{\text{0}},f_{{\text{0}},\text{gs}})\rangle$ and $(|\Phi_{\text{2}},f_{{\text{2}},\text{gs}})\rangle$) in different vison sectors into one of connecting two states ($|(\Phi_{\text{0}},f_{{\text{0}},\text{gs}})\rangle$ and $|(\Phi_{\text{0}},f_{{\text{0}},\text{gs}}^{\prime})\rangle$) in the same vison sector.
The remaining steps are similar to those used in GS preparation.

First, we consider the case of two Hamiltonians allowing identical physical fermion parity.
Then, we can determine the matrix $B$ as follows,
\begin{equation}
e^{B}=Q_{\text{0}}^{\prime}Q_{\text{0}}^{T}\quad \text{(identical parity)}.
\label{Eq:B_identical_parity_Vison}
\end{equation}
Under the action of the $U_{1}$ operator, $H(A_{\text{0}})$ transforms into $H(A_{\text{0}}^{\prime\prime})$ ($A_{\text{0}}^{\prime\prime}=Q_{\text{0}}^{\prime}E_{\text{0}}{Q_{\text{0}}^{\prime T}}$). As a result, the $|(\Phi_{\text{0}},f_{0,\text{gs}})\rangle$ is transformed to $|(\Phi_{\text{0}},f_{{\text{0}},\text{gs}}^{\prime})\rangle$. 
The $U_{1}$ operator gives the following rotation in fermion space,
\begin{equation}
\label{Eq:fermion_mode_rotation_equal_vison}
(a_{\text{0},1},a_{\text{0},2},\dotsc,a_{\text{0},N/2})\to(a_{\text{0}^{\prime},1},a_{\text{0}^{\prime},2},\dotsc,a_{\text{0}^{\prime},N/2}).
\end{equation}
Consequently, the total unitary operator $U_{2}U_{1}$ maps $|(\Phi_{\text{0}},f_{{\text{0}},\text{gs}})\rangle$ to $|(\Phi_{\text{2}},f_{{\text{2}},\text{gs}})\rangle$. 

Second, we need a modification if the initial and final vison sectors allow different physical fermion parity.
One can resolve this issue by adding the $T_{1,2}$ between $Q_{\text{0}}^{\prime}$ and $Q_{\text{0}}^{T}$
\begin{equation}
e^{B}=Q_{\text{0}}^{\prime}T_{1,2}Q_{\text{0}}^{T}\quad \text{(different parity)}.
\label{Eq:B_different_parity_Vison}
\end{equation}
The $U_{1}$ operator gives the following rotation in fermion space,
\begin{equation}
\label{Eq:fermion_mode_rotation_dif_vison}
(a_{\text{0},1},a_{\text{0},2},\dotsc,a_{\text{0},N/2})\to(a_{\text{0}^{\prime},1}^{\dagger},a_{\text{0}^{\prime},2},\dotsc,a_{\text{0}^{\prime},N/2}).
\end{equation}

An important note is that one can not use this vison manipulation process to perform braiding operations in topological quantum computation. 
In other words, this process can not produce the non-Abelian statistics as in computation works \cite{Lahtinen_2009,Bolukbasi_2012}. 
Unlike the usual braiding process, this process can assign the arbitrary phase to each fermion mode. 

The matrix $Q$ is obtained from the decomposition of the skew-symmetric matrix $A$ \eqref{Eq:decomposition}, which is not uniquely determined. 
Consider the following transformation acting on the matrix $Q$.
\begin{equation}
\label{Eq:Fermion_mode_U1_transformation}
    Q^{\prime}=Q\begin{bmatrix}
    +\text{cos}(\theta_{1}) & -\text{sin}(\theta_{1}) & &&\\
    +\text{sin}(\theta_{1}) & +\text{cos}(\theta_{1})&&&\\
    &&+\text{cos}(\theta_{2}) & -\text{sin}(\theta_{2}) & \\
    &&+\text{sin}(\theta_{2}) & +\text{cos}(\theta_{2})&\\
    & & &&\ddots\\
   \end{bmatrix}.
\end{equation}
This corresponds to local $U(1)$ transformation acting on each fermion mode ($a_{k}\to e^{+i\theta_{k}}a_{k}$ and $a_{k}^{\dagger}\to e^{-i\theta_{k}}a_{k}^{\dagger}$).
Acting this transformation on $Q_{\text{0}}^{\prime}$ will modify the mapping \eqref{Eq:fermion_mode_rotation_equal_vison} as follows,
\begin{gather*}
\label{Eq:fermion_mode_rotation_equal_vison_gauge}
(a_{\text{0},1},a_{\text{0},2},\dotsc)\to(e^{+i\theta_{1}}a_{\text{0}^{\prime},1},e^{+i\theta_{2}}a_{\text{0}^{\prime},2},\dotsc).
\end{gather*}

Physically, the usual braiding process, which involves adiabatic transport of vison, gives each fermion mode an arbitrary dynamical phase. 
If the trajectory forms a closed loop, one can obtain the topological phase while ignoring the effect of the dynamical phase.
On the other hand, in our theory, one can assign the $U(1)$ phase to each fermion mode in the $U_{1}$ operator. 
While this process is unsuitable for performing braiding operations, it can access the Majorana fermion excitations without a braiding operation.

\begin{figure*}[t]
\setcounter{figure}{9}
\includegraphics[width=1.0\linewidth]{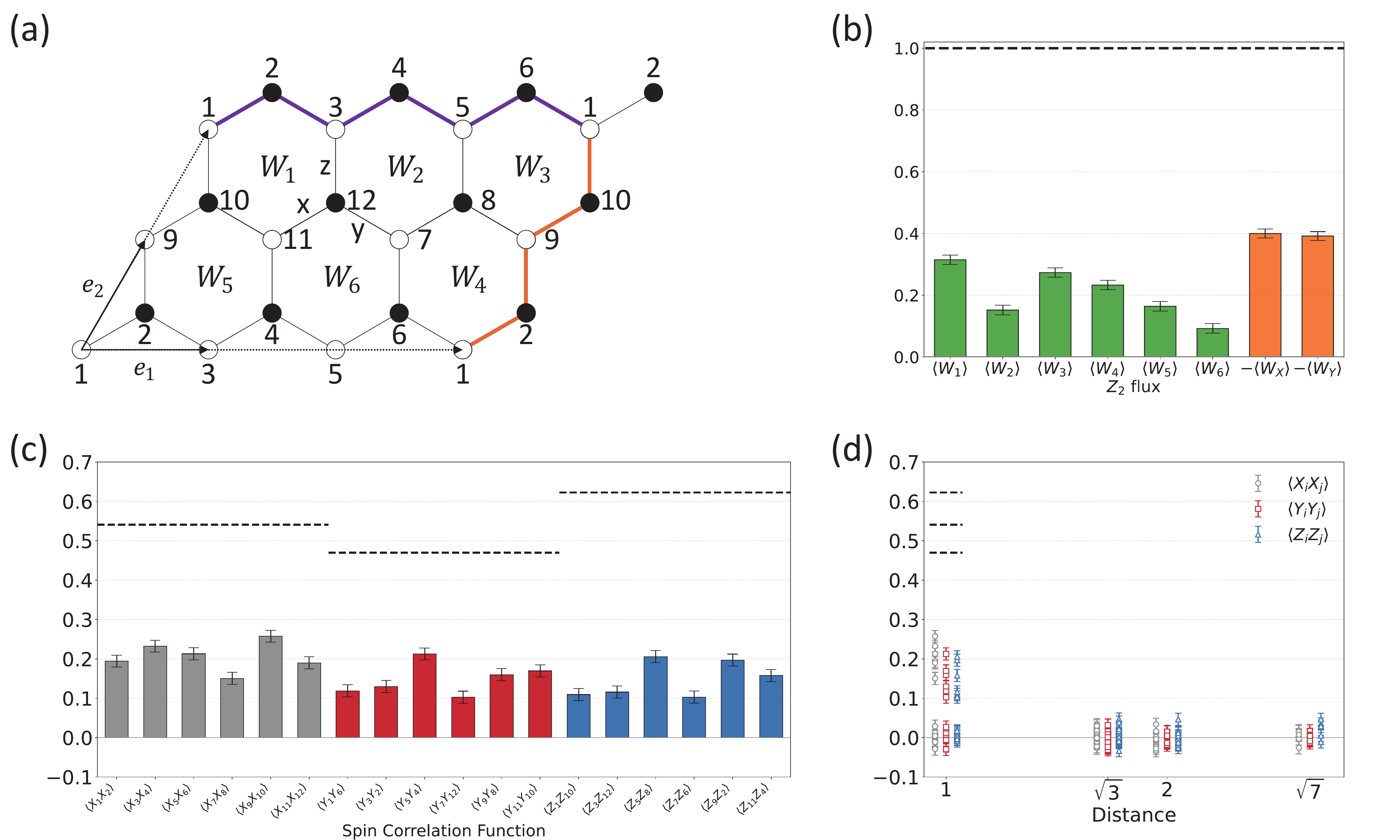}
\caption{Ground state preparation (12-qubit model). (a) Geometry of KQSL on the torus ($L_{1}=3$, $L_{2}=2$, and $M=0$). 
Dashed arrow connects identical sites. 
Purple (orange) loop indicates non-contractible loop $W_{X}$ ($W_{Y}$) on the torus.
(b) The expectation value of six $Z_{2}$ vison operator (green) and two Wilson loop operators (orange) (4096 shots in total).
(c) The spin correlation obtained from the data set. 
The spin correlation ($\langle\sigma^{\alpha}_{i}\sigma^{\alpha}_{j}\rangle$) is measured for 18 links connecting the pair of nearest neighbors on the torus, with specific $\alpha$. 
The dashed line indicates the spin correlation value obtained from theory.
The color (gray, red, and blue) indicates the $\alpha$=$x$, $y$, and $z$, respectively.
(d) The measured spin correlation function as a function of distance between two sites. 
The spin correlation function is obtained for all 66 pairs on the torus.
The color (gray, red, and blue) indicates the $\alpha$=$x$, $y$, and $z$, respectively.
}
\end{figure*}
\subsection{Majorana fermion control}
The $U_{1}$ operator allows us to access the Majorana fermion excitations without braiding operations, which corresponds to a gate operation in logical space. 
First, we define the two-dimensional logical space from the eigenstates of the original Hamiltonian. 
We select the ground and first excited state from a certain vison sector. 
The only requirement is that physical fermion parity be well-defined ($0<\epsilon_{1}\leq\epsilon_{2}\leq\dotsc\leq\epsilon_{N/2}$).
For a given vison sector, the logical space is determined by the corresponding physical fermion parity
\begin{gather*}
\begin{aligned}
&\{|(\Phi_{\text{i}},f_{{\text{i}},\text{vac}})\rangle,\,\,
a_{\text{i},2}^{\dagger}a_{\text{i},1}^{\dagger}|(\Phi_{\text{i}},f_{{\text{i}},\text{vac}})\rangle\}\quad\text{(even)}\\
&\{a_{\text{i},1}^{\dagger}|(\Phi_{\text{i}},f_{{\text{i}},\text{vac}})\rangle,\,\,
a_{\text{i},2}^{\dagger}|(\Phi_{\text{i}},f_{{\text{i}},\text{vac}})\rangle\}\quad\text{(odd)},
\end{aligned}
\end{gather*}
where $a_{\text{i},k}^{\dagger}$ is the fermionic creation operator obtained from diagonalizing quadratic Hamiltonian.
As an example, consider the unitary rotation that swaps the first and second fermionic modes
\begin{equation}
\label{Eq:fermion_mode_rotation}
(a_{\text{i},1},a_{\text{i},2},\dotsc,a_{\text{i},N/2})\to(a_{\text{i},2},a_{\text{i},1},\dotsc,a_{\text{i},N/2}).
\end{equation}

The $U_{1}$ operator can implement this rotation with the following skew-symmetric matrix $B$,
\begin{equation}
e^{B}=Q_{\text{i}}T_{1,3}T_{2,4} Q_{\text{i}}^{T}.
\label{Eq:logic_gate}
\end{equation}
$T_{1,3}$ ($T_{2,4}$) is the elementary row operation that swaps the first (second) and third (fourth) row vectors. 
If the allowed physical fermion parity is even, the $U_{1}$ acts as $\sigma^{z}$ in logical space (up to a global phase). 
On the other hand, if the parity is odd, the $U_{1}$ acts as $\sigma^{x}$ in logical space (up to a global phase).
We have demonstrated the simplest logical operations, but one can generalize this method to realize any logical operation that can be associated with the $U_1$ operator.

The $U_1$ operator can control the unpaired Majorana modes carried by visons. 
The energy of unpaired Majorana modes converges to zero as the distance between visons increases, but it is not exactly zero.
The energy shows exponential convergence as the vison separation increases \cite{Lahtinen_2011}. 
Thus, we can define the physical fermion parity with an energy mode close to 0. 
Typically, in our calculation, we can determine the physical fermion parity as long as the fermion energy is larger than $10^{-6}J$.

While the visons are characterized by gauge invariant operator $W_{p}$, fermionic excitations are generally impossible to measure with gauge invariant operator. 
However, we can resolve this issue by introducing the $U_{1}$ operator gives the mapping from eigenstates of $H_{\text{K}}$ to eigenstates of $H_{\text{dim}}$; this is the inverse process of GS preparation.

\begin{figure*}[t]
\setcounter{figure}{10}
\includegraphics[width=1.0\linewidth]{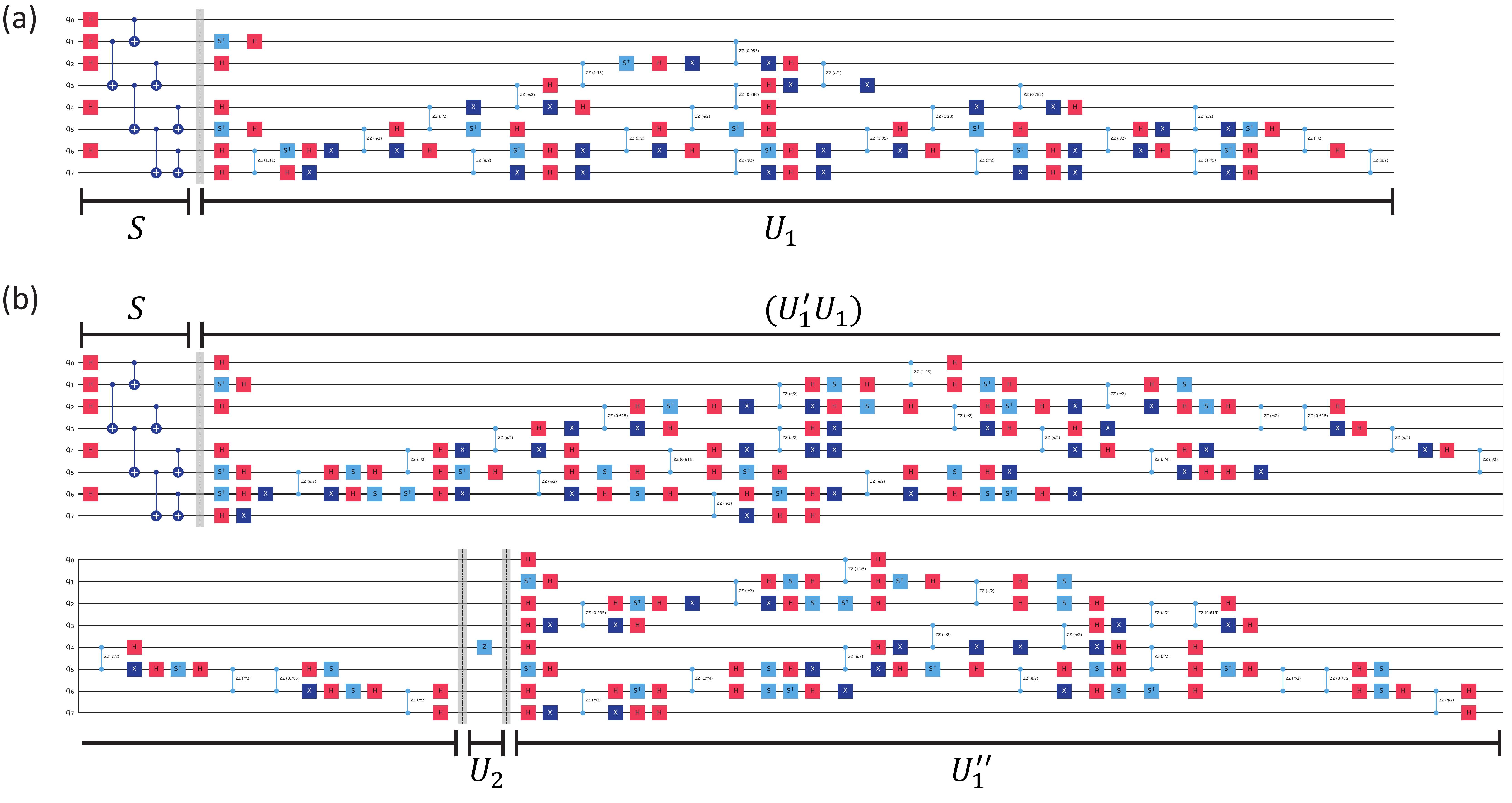}
\caption{(a) The quantum circuit designed for eight-qubit ground state preparation.
We use 21 $R$ operations to perform the $U_{1}$ operator for GS preparation.
(b) The quantum circuit designed for eight-qubit vison manipulation and Majorana fermion control.
For the vison manipulation experiment, the $U_{1}^{\prime\prime}$ operator is omitted.
We use 20 (14) $R$ operations to perform the $U_{1}^{\prime}U_{1}$ ($U_{1}^{\prime\prime}$) operator.
}
\end{figure*}
\subsection{Fermion readout}

We will introduce the explicit procedures to read fermionic excitations. 
Unlike the $H_{\text{K}}$, $H_{\text{dim}}$ \eqref{Eq:H_dim} has a direct relation between the fermionic excitation and local spin correlation function.
Thus, we construct the unitary operator to map the eigenstates of $H_{\text{K}}$ to the eigenstates of $H_{\text{dim}}$. 
One can understand it as an inverse process of GS preparation.

To relate the fermionic excitation and spin correlation function, we obtain the following quadratic form of $H_{\text{dim}}$ with $Z_2$ gauge field characterized by $\Phi_{k}$. 
For simplicity, we assign $J$ in ascending order ($0<J_{1}<J_{2}<\cdots<J_{N/2}$).
\begin{equation}
\begin{aligned}
H(A_{\text{dim}})&=\frac{i}{4}\sum_{i}2 J_{i}u_{2i-1,2i}(c_{2i-1}c_{2i}-c_{2i}c_{2i-1})\\
&=\sum_{i}2 J_{i}(a_{i}^{^{\dagger}}a_{i}-\frac{1}{2}).\\
\end{aligned}
\label{Eq:H_Quadartic_dim}
\end{equation}
Here, the two spins ($(2i-1)$th and $(2i)$th spin) are coupled by the Kitaev interaction with the $X$-direction.
We have $a_{i}^{\dagger}=\frac{1}{2}({c}_{2i-1}-i u_{2i-1,2i}{c}_{2i})$ and $a_{i}=\frac{1}{2}({c}_{2i-1}+i u_{2i-1,2i}{c}_{2i})$. 
From this relation, one can relate the fermionic excitation of $H(A_{\text{dim}})$ and the spin correlation function $\langle\sigma_{2i-1}^{x}\sigma_{2i}^{x}\rangle$. 
\begin{equation}
    \langle{}n_{i}\rangle{}=\frac{1}{2}(1-\langle{}\sigma_{2i-1}^{x}\sigma_{2i}^{x}\rangle{}).
\end{equation}
If the $i$th fermion mode is occupied ($\langle{}n_{i}\rangle{}=1$), $\langle{}\sigma_{2i-1}^{x}\sigma_{2i}^{x}\rangle{}=-1$. 
This relation is $Z_2$ gauge invariant: we impose that every fermion mode has positive energy to define the physical fermion.

We aim to measure the fermion excitations of $|(\Phi_{k},f_{k})\rangle{}$, which is a linear superposition of eigenstates of $H_{\text{K}}$ with the same vison configuration $\Phi_{k}$. 
Through the $U_{1}$ operator that maps the eigenstates of $H_{\text{K}}$ to the eigenstates of $H_{\text{dim}}$, one can map the state $|(\Phi_{k},f_{k})\rangle{}$ to $|(\Phi_{k},f_{k}^{\text{dim}})\rangle{}$. 
Thus, we can read the fermion excitations of $|(\Phi_{k},f_{k})\rangle{}$ as follows.
\begin{equation}
\begin{aligned}
\langle{}n_{i}\rangle{}&=\frac{1}{2}-\frac{1}{2}\langle{}(\Phi_{k},f_{k}^{\text{dim}})|\sigma_{2i-1}^{x}\sigma_{2i}^{x}|(\Phi_{k},f_{k}^{\text{dim}})\rangle{}\\
&=\frac{1}{2}-\frac{1}{2}\langle{}(\Phi_{k},f_{k})|U_{1}^{\dagger}\sigma_{2i-1}^{x}\sigma_{2i}^{x}U_{1}|(\Phi_{k},f_{k})\rangle{}.
\end{aligned}
\end{equation}
If two Hamiltonians $H_{\text{K}}$ and $H_{\text{dim}}$ allow different physical fermion parity, the relation will be modified; every even parity mode maps to odd parity mode and vice versa.

The $U_1$ operator, connecting eigenstates of $H_{\text{K}}$ and $H_{\text{dim}}$, acts as an encoder/decoder in the GS preparation/fermion readout process. 
In GS preparation, we construct the unitary operator that maps the eigenstates of $H_{\text{dim}}$ to the eigenstates of $H_{\text{K}}$ (encoding).
On the other hand, in the fermion readout, we construct the unitary operator that maps the eigenstates of $H_{\text{K}}$ to the eigenstates of $H_{\text{dim}}$ (decoding).
\begin{figure*}[t]
\setcounter{figure}{11}
\includegraphics[width=1.0\linewidth]{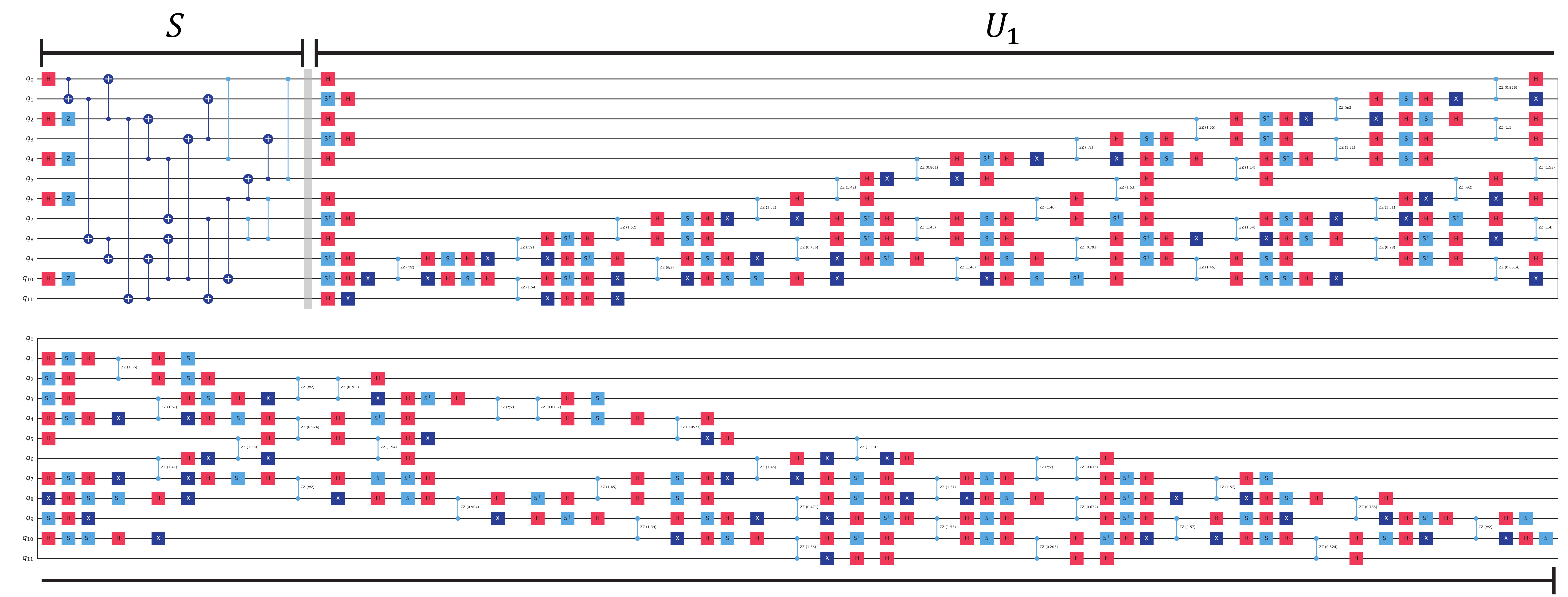}
\caption{The quantum circuit designed for 12-qubit ground state preparation.
We use 56 $R$ operations to perform the $U_{1}$ operator for GS preparation.
}
\end{figure*}

Although we constructed the unitary operator for the Majorana fermion readout, we could not test it experimentally.
The main reason is that the quantum circuit optimization tool may merge the unitary operators in a way that oversimplifies the circuit structure.
Circuit optimization may disregard the intended encoding/decoding structure, potentially allowing direct logical operations between local fermion modes.
If the quantum circuit is overly simplified through circuit optimization, it leaves room for controversy in interpreting the results as a genuine readout of fermionic excitations of the KQSL Hamiltonian.

\subsection{Implementation of $U_{1}$ operator}

This section will explicitly show the process of decomposing the $U_{1}$ operator to a set of local gate operations. 
First, we assign the index to each site using the following rules: (i) site on A (B) sublattice has an odd (even) index, (ii) if two sites have indexes that differ by 1, they are nearest neighbors. 
One can always find such indexing for KQSL model on the torus. 
Our goal is decomposing the $U_{1}$ operator to a product of $R(n_{i},\theta_{i})$ operators.
\begin{equation}
\label{Eq:Q_decomposition_R}
U_{1}=R(n_{M},\theta_{M})\times\dotsc\times{}R(n_{2},\theta_{2})\times{}R(n_{1},\theta_{1}).
\end{equation}
Each $R(n_{i},\theta_{i})$ operator corresponds to a local two-qubit operation. 
$\alpha=\text{$x$, $y$ or $z$}$ depending on the direction of the link.
\begin{equation}
\label{Eq:R_operation}
R(n,\theta)=e^{-\frac{i}{2}\theta{}u_{n,n+1}\sigma_{n}^{\alpha}\sigma_{n+1}^{\alpha}}.
\end{equation}
As discussed in previous sections, the special orthogonal matrix $e^{+B}$ can fully characterize the $U_{1}$ operator.
Now, we can rewrite the Eq. \eqref{Eq:Q_decomposition_R} in the fermion space.
\begin{equation}
\label{Eq:Q_decomposition_R_fermion}
e^{+B}=r(n_{M},\theta_{M})\times\dotsc\times{}r(n_{2},\theta_{2})\times{}r(n_{1},\theta_{1}).
\end{equation}
$r(n,\theta)$ is a special orthogonal matrix that mixes the fermion $c_{n}$ and $c_{n+1}$.
In fermion space, $r(n,\theta)$ is a Givens rotation that mixes the nearest rows.
\begin{equation}
r(n,\theta)=\begin{bmatrix}
    1 &  &   &&&\\
     & \ddots& & &&  \\
    & &+\text{cos}(\theta)  & -\text{sin}(\theta)&& \\
    & & +\text{sin}(\theta) & +\text{cos}(\theta)& &\\
    & & &  & \ddots  &\\
    & & &  &   &1\\
   \end{bmatrix}.
\end{equation}

We will provide an alternative viewpoint to interpret this problem. 
One can view this problem of decomposing matrix $e^{+B}$ as a problem finding the inverse matrix of $e^{-B}$ as a product of the $r$ matrix. 
Then, we can use the basic techniques to find the inverse matrix. 
In standard Gaussian elimination, to find the inverse of an invertible matrix, one successively applies elementary row operations to find a reduced row echelon form (RREF) of a matrix.
Instead of elementary row operations, we use the $r$ matrix (Givens rotation) to eliminate components below the diagonal. 
The matrix becomes row echelon form (REF) after eliminating all components below the diagonal. 
At every step, the matrix remains special orthogonal; thus, the REF is a diagonal matrix that has an even number of $-1$ in diagonal and else is $+1$. 
Again, $-1$s in diagonal can be flipped into $+1$ by successively applying the $r$ matrix ($\pi$ rotation).
The $\pi$ rotation has a special property; $r(n,\pi)\,r(n+1,\theta)=r(n+1,-\theta)\,r(n,\pi)$.
These $\pi$ rotations can be absorbed into other $r$ operations from this relation. 
The overall procedure is illustrated in Fig. 9(a). 

As the $U_{1}$ operator does not affect the vison degree of freedom, it can be fully characterized by a special orthogonal matrix $e^{+B}$.
While the $U_{1}\in{}U(2^{N})$ has an exponentially increasing dimension, the $e^{+B}\in{}SO(N)$ has a linearly increasing dimension, which simplifies the decomposition problem.
The strategy of decomposing the unitary operator into a sequence of local two-qubit gates (corresponding to Givens rotations) has been explored in the studies to simulate the quadratic Hamiltonians in a more general context \cite{Wecker_2015,Kivlichan_2018,Jiang_2018}.
Our approach is specifically adapted to resolve the particular problems arising from the KQSL Hamiltonian.
After we obtain the $r$ operations decomposing $e^{+B}$, $r$ operations (in the fermion space) can be converted to $R$ operations in the qubit space.
$R$ operations can be implemented with the following quantum gate operations illustrated in Fig. 9(b) and 9(c).

In summary, one can decompose the $U_{1}$ operator into a sequential product of the $R$ operators.
For $N$ spin model, $U_{1}$ can be decomposed into $N(N-1)/2$ $R$ operations. 
This decomposition allows a local quantum circuit within the depth $(2N-3)$ to perform the $U_{1}$ operator. 
While the matrix $B$ is not gauge invariant, the $R$ operation in the qubit space is $Z_{2}$ gauge invariant. 
In the Eq. \eqref{Eq:R_operation}, $\theta$ and $u_{n,n+1}$ are not gauge invariant, but $(\theta u_{n,n+1})$ is a gauge invariant quantity. 

The correspondence between $U_{1}$ operator and $e^{+B}$ makes it easy to show that $U_{1}$ is closed under multiplication. 
As $e^{+B}$ belongs to $SO(N)$, $e^{+B}$ is closed under multiplication.
Consequently, the single $U_{1}$ operator can emulate the sequential application of time evolution generated by different quadratic Hamiltonians that preserves vison configuration, with the quantum circuit depth proportional to the system size.
As the circuit depth does not depend on the time scale (or the number of different quadratic Hamiltonians), it will be advantageous to simulate long-time dynamics of the KQSL model.
A typical example we are interested in is studies \cite{Evered_2025,Will_2025} that explored the dynamics of the periodically driven KQSL model. 
In these studies, one can interpret the Floquet drive as a sequential application of time evolution generated by different quadratic Hamiltonians that can be replaced by a single $U_{1}$ operator.

\section{Ground state preparation: 12-qubit model}
We test our ground state preparation process to the 12-qubit KQSL model ($K/J=0$), as illustrated in Fig. 10(a). 
Although the experimental noise becomes more significant in the 12-qubit model, the experimental data still captures the key properties of the KQSL ground state.
In this model, the ground state lies in the vison-free sector. 
The explicit construction of the full quantum circuit for this model is provided in Appendix D.

Figures 10(c) and 10(d) present the measured spin correlation functions, while the expectation value of the vison and Willson loop operators are shown in Fig. 10(b).
The measured energy expectation value is,
\[{\langle{}E\rangle}_{\text{exp}}=-3.0180 \,(\pm0.2769) \,J.\] 
While the exact value is
\[{\langle{}E\rangle}_{\text{exact}}=-9.8002\,J.\]
Compared to the eight-qubit GS preparation, the measured energy expectation value significantly differs from the exact value.
Due to the increased experimental noise, we could not further implement vison manipulation or Majorana fermion control in the 12-qubit model.

\section{Quantum Circuit Details}
This section will discuss the theoretical and technical details for implementing the designed quantum circuit in actual quantum processors.
The full quantum circuits used for eight-qubit (12-qubit) KQSL simulation is illustrated in Fig. 11 (Fig. 12).
These circuits can be implemented in an actual experiment after performing transpiliation (with quantum circuit optimization).

A few important notes are following.
First, with the proper $U(1)$ gauge (related to local $U(1)$ transformation \eqref{Eq:Fermion_mode_U1_transformation}), we reduce the number of $R$ operations by using proper criteria.
For example, $R(n, \theta)$ operations are excluded from the quantum circuit if $|\theta|<10^{-6}$, which leads to the quantum circuit could be constructed with a reduced number of $R$ operations.
Second, in the Fig. 11(b), we combine the $U_{1}$ and $U_{1}^{\prime}$ operators to reduce the number of $R$ operations.
As we discussed in Appendix B, the fermion rotation operators ($U_{1}$ and $U_{1}^{\prime}$) are closed under multiplication.
Third, we use the RZZ gate to replace two CX gates and one RZ gate in Fig 9 (c).
As our quantum processor, IBM Heron r2 processor 'ibm-marrakesh', supports the RZZ gate as a basis gate, it can implement the $R$ operation efficiently.

\begin{figure}[t]
\setcounter{figure}{12}
\includegraphics[width=0.90\linewidth]{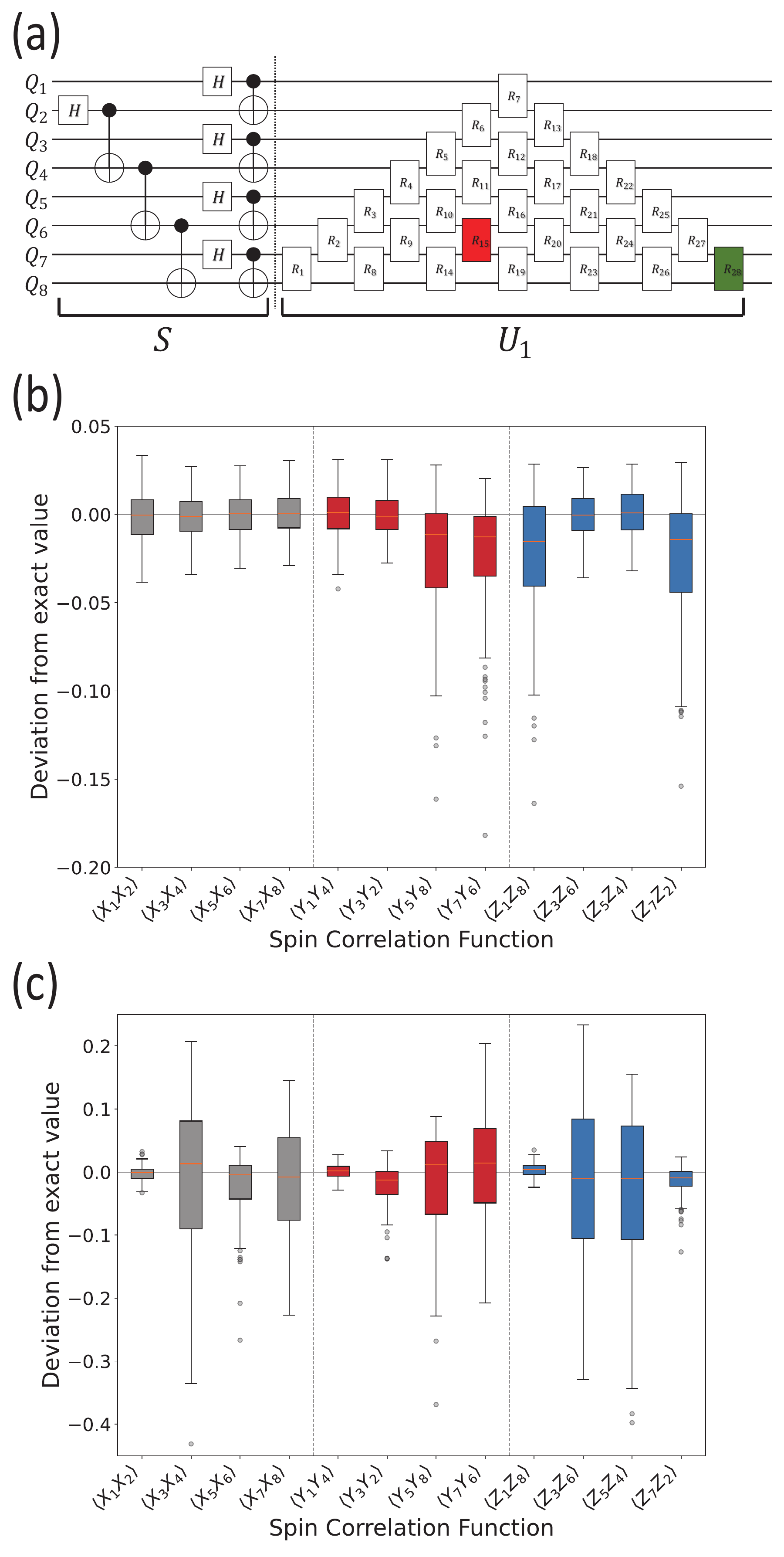}
\caption{(a) Schematic diagram of the quantum circuit for eight-qubit GS preparation.
We introduce the Gaussian error to the 28th (15th) $R$ block, which is indicated by green (red) color; $\delta\theta\sim{}N(0,\Delta^{2})$, with $\Delta=0.3\text{ rad}$.
(b) and (c) The spin correlation obtained from 'numerical' simulation, with the Gaussian error introduced to the 28th (15th) $R$ block.
We sample the $\delta\theta$ 100 times and make 4096 measurements for each $\delta\theta$ to obtain the correlation function.
The spin correlation ($\langle\sigma^{\alpha}_{i}\sigma^{\alpha}_{j}\rangle$) is measured for 12 links connecting the pair of nearest neighbors on the torus, with specific $\alpha$. 
The color (gray, red, and blue) indicates the $\alpha$=$x$, $y$, and $z$, respectively.
The vertical axis represents the deviation from the ideal value.}
\end{figure}
\section{Error Analysis}
Here, we provide a detailed analysis of the experimental error sources and their impact on the physical observables.
We use the IBM Heron r2 processor 'ibm-marrakesh' to perform the experiment. 
This device features an average two-qubit gate (CZ gate and RZZ gate) error rate of approximately $\sim{}10^{-3}$, a single-qubit gate error rate of $\sim{}10^{-4}$, and a readout error of $\sim{}10^{-2}$; one can find explicit calibration data on the IBM website \cite{IBM_marrakesh}.
According to the provided calibration data, the dominant source of infidelity is the accumulated two-qubit gate errors.
Although the readout error rate is higher than that of an individual two-qubit error rate, the cumulative effect makes the latter the dominant source of total infidelity.
To characterize the prepared KQSL eigenstate, we measure the expectation values of the $W_p$ operators (vison) and spin correlation functions.

First, as the $W_p$ operator is a six-qubit operator, it is highly susceptible to local errors. 
A single Pauli error occurring on the given plaquette flips the sign of the measured vison value and thus appears as a vison excitation error.
It poses a significant challenge as system size increases.
The $U_1$ operator, with a circuit depth of $O(N)$, is decomposed into $O(N^2)$ $R$ operations, resulting in a total RZZ gate count that scales as $O(N^2)$.
Note that one RZZ gate and single-qubit rotations are required to perform one $R$ operation \eqref{Eq:R_operation}.
While Fig. 9(c) shows the decomposition of the $R$ operation with two CX gates, we use RZZ gate to implement the $R$ operation in a real experiment.
For the ground state preparation of the eight-qubit system (Fig. 11(a)), we implement the $U_1$ operator using 21 $R$ operations. 
In contrast, the 12-qubit implementation (Fig. 12) requires 56 $R$ operations. 
\begin{gather*}
    F_{\text{vison}}\approx{}{(1-\epsilon_{\text{RZZ}})}^{N_{\text{RZZ}}} 
\end{gather*}
Using a simplified fidelity model, here $\epsilon_{\text{RZZ}}$ is the RZZ gate error rate; the increase in two-qubit gate count leads to an exponential decay in fidelity for vison sector identification. 
This is the primary reason for the larger deviation from theoretical predictions observed in the 12-qubit experiment.

Second, while the spin correlation function is intrinsically less susceptible to local errors, the structure of the quantum circuit makes it susceptible to local errors.
The implementation of the $U_{1}$ operator requires a strong condition that the vison sector remains fixed throughout the quantum circuit.
As we discussed, to simplify the problem, we restrict the $U_1$ operator to a specific vison sector; the $U_1$ operator is only valid within this specific vison sector.
As a result, the spin correlation function measurement shares similar vulnerabilities to the vison measurement.

Additionally, the structure of the $U_1$ operator is prone to error propagation. 
In constructing the quantum circuit for implementing the $U_1$ operator, we stack the $R$ operations in a pyramid-like structure while performing two-qubit gates on qubits with neighboring indices.
We design this structure to perform the basis transformation from one fermionic basis to another in a 1D system, as in Eq. \eqref{Eq:fermion_mode_rotation_equal}. 
This structure controls all fermionic degrees of freedom with a minimal number of gate operations, allowing errors to be easily propagated.
We will discuss an explicit example of how the spin correlation function is affected due to the error propagation.
Consider the following error model that affects the rotation angle of the RZZ gate.

\begin{gather*}
R_{ZZ}(\theta_{\text{noisy}})=R_{ZZ}(\theta_{\text{ideal}}+\delta\theta),\,\,\,\delta\theta\sim{}N(0,\Delta^{2})
\end{gather*}
We introduce the Gaussian error to the rotation angle obtained from the $U_1$ operator construction; this is a specific type of coherent error that only affects the fermionic degree and has no change in the vison configuration.
As it affects the fermionic degree, it changes the spin correlation function. 
By numerical calculation, we will demonstrate how an error (in one RZZ gate) propagates and affects the spin correlation function.
We revisit the eight-qubit GS preparation process, introduce the error, and calculate how the spin correlation function changes depending on where the error is located.
The Fig. 13(a) represents two cases in which errors are introduced in different locations.

First, we consider the case where the error is introduced to the 28th $R$ block, the last block. 
Obviously, as the error is introduced in the last block, the error cannot propagate through the quantum circuit.
As a result, only four correlation functions, containing either site 7 or 8, are affected by error, see Fig. 13(b).
Note that the correlation function $\langle{}X_7 X_8 \rangle$ is not affected by error, as the operator ($\sigma_7^{x} \sigma_8^{x}$) commutes with the error.

Second, we consider the case where the error is introduced to the 15th $R$ block, in the middle of the quantum circuit. 
In this case, the error can propagate through the quantum circuit.
As a consequence, more correlation functions are affected by errors, see Fig. 13(c).
Limiting the type of error makes it easier to analyze the error propagation.
As we discussed in Appendix B, the $U_1$ operator is closed under multiplication.
Thus, using the fermionic representation, one can decompose the 'noisy' $U_1$ operator as follows. 
\begin{equation}
    U_{1,\text{ noisy}}=U_{1,\text{ propagation}}\,U_{1,\text{ ideal}}
\end{equation}
Even when multiple errors are introduced, this decomposition is possible if the errors do not affect the vison sector and remain coherent.
While our analysis focuses on a single coherent error, in a real experiment, the final infidelity results from an interplay of multiple coherent and incoherent error sources. 
\clearpage
\bibliography{reference}

\end{document}